\documentclass[12pt]{iopart}

\usepackage[dvipdfm]{graphicx}

\begin{document}

\title[Distribution of Lee-Yang zeros in the $\pm J$ model]{Distribution of Lee-Yang zeros and Griffiths singularities in the $\pm J$ model of spin glasses}

\author{Yoshiki Matsuda$^{1}$, Hidetoshi Nishimori$^{1}$ and Koji Hukushima$^{2}$}

\address{$^{1}$Department of Physics, Tokyo Institute of Technology, Oh-okayama, Meguro-ku, Tokyo 152-8551, Japan}
\address{$^{2}$Department of Basic Science, Graduate School of Arts and Sciences, University of Tokyo, Komaba, Meguro-ku, Tokyo 153-8902, Japan}
\ead{matsuda@stat.phys.titech.ac.jp}

\begin{abstract}

We investigate the distribution of zeros of the partition function of the two- and three-dimensional symmetric $\pm J$ Ising spin glasses on the complex field plane. We use the method to analytically implement the idea of numerical transfer matrix which provides us with the exact expression of the partition function as a polynomial of fugacity. The results show that zeros are distributed in a wide region in the complex field plane. Nevertheless we observe that zeros on the imaginary axis play dominant roles in the critical behaviour since zeros on the imaginary axis are in closer proximity to the real axis. We estimate the density of zeros on the imaginary axis by an importance-sampling Monte Carlo algorithm, which enables us to sample very rare events. Our result suggests that the density has an essential singularity at the origin. This observation is consistent with the existence of Griffiths singularities in the present systems. This is the first evidence for Griffiths singularities in spin glass systems in equilibrium.

\end{abstract}

%\maketitle

\section{Introduction}
In order to shed a unique light on the problem of finite-dimensional spin glasses, we discuss the distribution of zeros of the partition function of the $\pm J$ Ising spin glass in the complex field plane, the Lee-Yang zeros \cite{LY}. All Lee-Yang zeros of a ferromagnetic system are proved to lie on the imaginary axis in the complex field plane from the circle theorem \cite{LY}. However, locations of Lee-Yang zeros for spin glass models in general do not obey simple rules. It is therefore interesting and important to study the distribution numerically in order to extract useful information on phase transitions in finite-dimensional spin glasses from a point of view different from conventional numerical methods like Monte Carlo simulations. The first study along this line was due to Ozeki and Nishimori \cite{Ozeki}. From data for relatively small systems, they found the tendency which may be consistent with the existence of the Griffiths singularity \cite{Griffiths}. Our present work is to expand their study to much larger systems and carry out quantitative evaluations of the density of zeros.

In the diluted ferromagnet, the Griffiths singularity means that the free energy is a non-analytic function of the field at the origin in the temperature range $T_c < T < T_c^{(\mathrm{pure})}$, where $T_c$ is the critical temperature and $T_c^{(\mathrm{pure})}$ is that of the corresponding non-random system. This temperature range is called the Griffiths phase. This unusual phenomenon in the otherwise-paramagnetic phase is caused by rare events of bond configurations, large-size connected clusters generated with small but non-vanishing probabilities. This fact was connected with an essential singularity in the density of zeros at the origin \cite{Bray} because large clusters tend to have zeros close to the origin at temperatures below $T_c^{(\mathrm{pure})}$. The Griffiths singularity in the diluted ferromagnet has also been studied through the behaviour of the inverse susceptibility analytically \cite{Bray-Moore} and numerically \cite{huku}.

On the spin glass problem, Randeria {\itshape et al.} studied the dynamics and suggested the existence of the Griffiths phase also in the spin glass systems \cite{Randeria}. There are few studies of Lee-Yang zeros of spin glass models in equilibrium except the work by Ozeki and Nishimori \cite{Ozeki} although there are some papers \cite{Bhanot, Damgaard} on the distribution of zeros in the complex temperature plane, Fisher zeros \cite{Fisher}. In the present paper we evaluate the density of Lee-Yang zeros of the spin glass systems by an importance-sampling Monte Carlo algorithm \cite{huku, Hartmann, Korner, Monthus, HandI}. Our results give evidence that the density of zeros has an essential singularity, which suggests the existence of the Griffiths singularity in spin glasses. 

The outline of this paper is as follows. In section 2 we explain the exact evaluation of the partition function by numerical transfer matrix. Then, we show the distribution of zeros of the $\pm J$ Ising model on the complex field plane in section 3. In section 4, we analyze zeros that have direct relevance to the phase transition and compare our investigation with known characteristics of phase diagrams. In section 5, we examine the density of zeros using the importance-sampling Monte Carlo algorithm which enables us to sample very rare events, and we discuss the existence of the Griffiths singularity in the $\pm J$ Ising model. We present our conclusion in section 6.

\section{Exact partition function}

We consider the $\pm J$ Ising model in an external magnetic field $h$ on the square and the simple cubic lattices with cylindrical boundary conditions, free in one direction and periodic in the others, which are suited for the transfer matrix method because we take the trace of spins one by one in spiral order. The Hamiltonian is
\begin{equation}
\mathcal{H} =  - \sum\limits_{\left\langle {i,j} \right\rangle } {J_{ij} \sigma _i \sigma _j }  - h\sum\limits_i {\sigma _i } ,
\end{equation}
where $\left\langle {i,j} \right\rangle$ denotes nearest neighbour pairs and $\sigma _i  =  \pm 1$. The interaction $J_{ij}$ is chosen as $J\left( { > 0} \right)$ or $-J$ with probability $1/2$. Using the number of states $\Omega \left( {E,M} \right)$ for given $E \equiv \frac{1}{2}\sum\nolimits_{\left\langle {i,j} \right\rangle } {\left( {1 - J_{ij} \sigma _i \sigma _j } \right)} $ and $M \equiv \frac{1}{2}\sum\nolimits_i {\left( {1 + \sigma _i } \right)} $, the partition function is expressed in terms of a two-variable polynomial as
\begin{eqnarray}
 Z\left(x,y\right) &= {\mathop{\Tr}\nolimits} \exp \left( { - \beta \mathcal{H}} \right) \\   &= y^{ - N_{s}/2} x^{ - N_{b}/2} \sum\limits_{M = 0}^{N_{s}} {\sum\limits_{E = 0}^{N_{b}} {\Omega \left( {E,M} \right)x^E } y^M } ,
\end{eqnarray}
where $y \equiv \exp \left( {2\beta h} \right)$ (fugacity), $x \equiv \exp \left( {2\beta J} \right),$ $N_s$ is the number of sites (assumed to be even), and $N_{b}$ is the number of bonds on the lattice.

We focus on the $y$-dependence of $Z\left(x,y\right)$ for a fixed $x$, since we are interested only in the field term for the investigation of Lee-Yang zeros. Thus, we treat the partition function as a single-variable polynomial 
\begin{equation}
Z\left( y \right) \equiv \sum\limits_{M = 0}^{N_s } {c_M y^M } \qquad \left(c_M  = \sum\limits_{E = 0}^{N_b } {\Omega \left( {E,M} \right)x^E } \right).
\end{equation}
To locate Lee-Yang zeros for the present model, we have to evaluate the partition function for a given set of random interactions analytically and solve the equation $Z(y)=0$ for $y$.

There are three methods to evaluate the partition function at least partially analytically using the idea of numerical transfer matrix \cite{Binder,IMKB}.

The first one is to evaluate the partition function at $N_s+1$ different external fields $y_i = \exp \left( 2\beta h_i\right)$ $\left(i=1,2,\cdots,N_s+1\right)$ by using the numerical transfer matrix method. Then, we solve the set of linear equations $\sum\nolimits_{M = 0}^{N_s } {c_M y_i^M }  = Z \left( y_i \right)$ $\left(i=1,2,\cdots,N_s+1\right)$ for $c_M $ $\left(M=0,1,\cdots,N_s\right)$ \cite{Ozeki}. In this way we know the values of all coefficients of the polynomial $Z \left( y \right) = \sum\nolimits_M {c_M y^M }$ and therefore we can solve $Z(y)=0$. In this process, since both $x$ and $y$ are treated as numbers, we have to be careful to keep very high precisions to avoid rounding errors in solving $Z(y)=0$. In addition, this method is not very efficient because it requires a repetition of essentially the same calculations. An advantage is that the required memory size is very small and is easily parallelized on the grid computer. We use this technique for systems larger than $16 \times 16$ (see the Appendix).

The second method, called the canonical transfer matrix, is to numerically give analytic partition functions at a fixed temperature for all values of $M$~\cite{Creswick}. This method evaluates $c_M$ (written as $\Omega \left( M, y \right)$ in \cite{Creswick}) for all $M$ at once. The required memory size is $N_s+1$ times larger than in the first method, but this process is done with only one calculation. Thus, the second method is faster than the previous method on a single CPU, but attention needs to be paid to the precision since we treat $x$ as a numerical value.

The third method is called the microcanonical transfer matrix \cite{Creswick}, which evaluates perfectly exact partition functions for all values of $M$ and $E$ using symbolic manipulations on the computer. This algorithm does not involve numerical errors. However, the memory requirement is $N_b$ times larger than in the second method.

In consideration of the balance between the computational time and the required memory size, we mainly use the canonical transfer matrix method in the present work. This method accompanies rounding errors. The necessary precision depends on the ratio of the largest and the smallest coefficients, which increases very rapidly with the system size or with the temperature decrease. We have verified the reliability of this approach for the ferromagnetic Ising model of sizes comparable to the spin glass system we are interested in as explained in the Appendix.

The system size we calculated is up to $20 \times 20$ in two dimensions and $4 \times 5 \times 8$ in three dimensions, where the last digit is for the free boundary direction and the others are for the periodic boundary conditions. We investigated at least about $10,000$ samples of random interactions for each system size. We solved the polynomial equation $Z(y)=0$ by \textsc{mathematica}$^{\rm{TM}}$ by specifying the precision to appropriate values.

\section{Distribution of zeros}

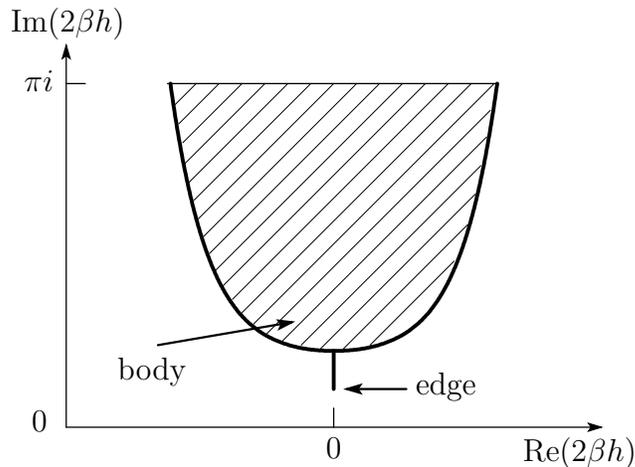
\begin{figure}
\begin{minipage}{\textwidth}
\begin{center}
%WinTpicVersion3.08
\unitlength 0.1in
\begin{picture}( 32.9000, 22.5000)(  5.1000,-22.5000)
% STR 2 0 3 0
% 3 2200 2210 2200 2310 5 0
% $0$
\put(22.0000,-23.1000){\makebox(0,0){$0$}}%
% FUNC 0 0 3 0
% 10 800 400 3800 2200 2200 2200 2400 2200 2200 2000 800 400 3800 2200 0 3 0 0 0 0
% (Exp(x)+Exp(-x))/10+1.8
\special{pn 20}%
\special{pa 1346 400}%
\special{pa 1346 402}%
\special{pa 1350 438}%
\special{pa 1356 472}%
\special{pa 1360 506}%
\special{pa 1366 540}%
\special{pa 1370 572}%
\special{pa 1376 602}%
\special{pa 1380 634}%
\special{pa 1386 664}%
\special{pa 1390 692}%
\special{pa 1396 720}%
\special{pa 1400 748}%
\special{pa 1406 776}%
\special{pa 1410 802}%
\special{pa 1416 828}%
\special{pa 1420 852}%
\special{pa 1426 876}%
\special{pa 1430 900}%
\special{pa 1436 924}%
\special{pa 1440 946}%
\special{pa 1446 968}%
\special{pa 1450 990}%
\special{pa 1456 1010}%
\special{pa 1460 1032}%
\special{pa 1466 1052}%
\special{pa 1470 1070}%
\special{pa 1476 1090}%
\special{pa 1480 1108}%
\special{pa 1486 1126}%
\special{pa 1490 1144}%
\special{pa 1496 1160}%
\special{pa 1500 1178}%
\special{pa 1506 1194}%
\special{pa 1510 1210}%
\special{pa 1516 1226}%
\special{pa 1520 1240}%
\special{pa 1526 1256}%
\special{pa 1530 1270}%
\special{pa 1536 1284}%
\special{pa 1540 1298}%
\special{pa 1546 1310}%
\special{pa 1550 1324}%
\special{pa 1556 1336}%
\special{pa 1560 1350}%
\special{pa 1566 1362}%
\special{pa 1570 1372}%
\special{pa 1576 1384}%
\special{pa 1580 1396}%
\special{pa 1586 1406}%
\special{pa 1590 1418}%
\special{pa 1596 1428}%
\special{pa 1600 1438}%
\special{pa 1606 1448}%
\special{pa 1610 1458}%
\special{pa 1616 1466}%
\special{pa 1620 1476}%
\special{pa 1626 1484}%
\special{pa 1630 1494}%
\special{pa 1636 1502}%
\special{pa 1640 1510}%
\special{pa 1646 1518}%
\special{pa 1650 1526}%
\special{pa 1656 1534}%
\special{pa 1660 1542}%
\special{pa 1666 1548}%
\special{pa 1670 1556}%
\special{pa 1676 1562}%
\special{pa 1680 1570}%
\special{pa 1686 1576}%
\special{pa 1690 1582}%
\special{pa 1696 1590}%
\special{pa 1700 1596}%
\special{pa 1706 1602}%
\special{pa 1710 1608}%
\special{pa 1716 1612}%
\special{pa 1720 1618}%
\special{pa 1726 1624}%
\special{pa 1730 1628}%
\special{pa 1736 1634}%
\special{pa 1740 1640}%
\special{pa 1746 1644}%
\special{pa 1750 1648}%
\special{pa 1756 1654}%
\special{pa 1760 1658}%
\special{pa 1766 1662}%
\special{pa 1770 1666}%
\special{pa 1776 1670}%
\special{pa 1780 1674}%
\special{pa 1786 1678}%
\special{pa 1790 1682}%
\special{pa 1796 1686}%
\special{pa 1800 1690}%
\special{pa 1806 1694}%
\special{pa 1810 1698}%
\special{pa 1816 1700}%
\special{pa 1820 1704}%
\special{pa 1826 1708}%
\special{pa 1830 1710}%
\special{pa 1836 1714}%
\special{pa 1840 1716}%
\special{pa 1846 1720}%
\special{pa 1850 1722}%
\special{pa 1856 1724}%
\special{pa 1860 1728}%
\special{pa 1866 1730}%
\special{pa 1870 1732}%
\special{pa 1876 1734}%
\special{pa 1880 1738}%
\special{pa 1886 1740}%
\special{pa 1890 1742}%
\special{pa 1896 1744}%
\special{pa 1900 1746}%
\special{pa 1906 1748}%
\special{pa 1910 1750}%
\special{pa 1916 1752}%
\special{pa 1920 1754}%
\special{pa 1926 1756}%
\special{pa 1930 1758}%
\special{pa 1936 1760}%
\special{pa 1940 1762}%
\special{pa 1946 1764}%
\special{pa 1950 1764}%
\special{pa 1956 1766}%
\special{pa 1960 1768}%
\special{pa 1966 1770}%
\special{pa 1970 1772}%
\special{pa 1976 1772}%
\special{pa 1980 1774}%
\special{pa 1986 1776}%
\special{pa 1990 1776}%
\special{pa 1996 1778}%
\special{pa 2000 1778}%
\special{pa 2006 1780}%
\special{pa 2010 1782}%
\special{pa 2016 1782}%
\special{pa 2020 1784}%
\special{pa 2026 1784}%
\special{pa 2030 1786}%
\special{pa 2036 1786}%
\special{pa 2040 1788}%
\special{pa 2046 1788}%
\special{pa 2050 1788}%
\special{pa 2056 1790}%
\special{pa 2060 1790}%
\special{pa 2066 1792}%
\special{pa 2070 1792}%
\special{pa 2076 1792}%
\special{pa 2080 1794}%
\special{pa 2086 1794}%
\special{pa 2090 1794}%
\special{pa 2096 1794}%
\special{pa 2100 1796}%
\special{pa 2106 1796}%
\special{pa 2110 1796}%
\special{pa 2116 1796}%
\special{pa 2120 1798}%
\special{pa 2126 1798}%
\special{pa 2130 1798}%
\special{pa 2136 1798}%
\special{pa 2140 1798}%
\special{pa 2146 1798}%
\special{pa 2150 1800}%
\special{pa 2156 1800}%
\special{pa 2160 1800}%
\special{pa 2166 1800}%
\special{pa 2170 1800}%
\special{pa 2176 1800}%
\special{pa 2180 1800}%
\special{pa 2186 1800}%
\special{pa 2190 1800}%
\special{pa 2196 1800}%
\special{pa 2200 1800}%
\special{pa 2206 1800}%
\special{pa 2210 1800}%
\special{pa 2216 1800}%
\special{pa 2220 1800}%
\special{pa 2226 1800}%
\special{pa 2230 1800}%
\special{pa 2236 1800}%
\special{pa 2240 1800}%
\special{pa 2246 1800}%
\special{pa 2250 1800}%
\special{pa 2256 1798}%
\special{pa 2260 1798}%
\special{pa 2266 1798}%
\special{pa 2270 1798}%
\special{pa 2276 1798}%
\special{pa 2280 1798}%
\special{pa 2286 1796}%
\special{pa 2290 1796}%
\special{pa 2296 1796}%
\special{pa 2300 1796}%
\special{pa 2306 1794}%
\special{pa 2310 1794}%
\special{pa 2316 1794}%
\special{pa 2320 1794}%
\special{pa 2326 1792}%
\special{pa 2330 1792}%
\special{pa 2336 1792}%
\special{pa 2340 1790}%
\special{pa 2346 1790}%
\special{pa 2350 1788}%
\special{pa 2356 1788}%
\special{pa 2360 1788}%
\special{pa 2366 1786}%
\special{pa 2370 1786}%
\special{pa 2376 1784}%
\special{pa 2380 1784}%
\special{pa 2386 1782}%
\special{pa 2390 1782}%
\special{pa 2396 1780}%
\special{pa 2400 1778}%
\special{pa 2406 1778}%
\special{pa 2410 1776}%
\special{pa 2416 1776}%
\special{pa 2420 1774}%
\special{pa 2426 1772}%
\special{pa 2430 1772}%
\special{pa 2436 1770}%
\special{pa 2440 1768}%
\special{pa 2446 1766}%
\special{pa 2450 1764}%
\special{pa 2456 1764}%
\special{pa 2460 1762}%
\special{pa 2466 1760}%
\special{pa 2470 1758}%
\special{pa 2476 1756}%
\special{pa 2480 1754}%
\special{pa 2486 1752}%
\special{pa 2490 1750}%
\special{pa 2496 1748}%
\special{pa 2500 1746}%
\special{pa 2506 1744}%
\special{pa 2510 1742}%
\special{pa 2516 1740}%
\special{pa 2520 1738}%
\special{pa 2526 1734}%
\special{pa 2530 1732}%
\special{pa 2536 1730}%
\special{pa 2540 1728}%
\special{pa 2546 1724}%
\special{pa 2550 1722}%
\special{pa 2556 1720}%
\special{pa 2560 1716}%
\special{pa 2566 1714}%
\special{pa 2570 1710}%
\special{pa 2576 1708}%
\special{pa 2580 1704}%
\special{pa 2586 1700}%
\special{pa 2590 1698}%
\special{pa 2596 1694}%
\special{pa 2600 1690}%
\special{pa 2606 1686}%
\special{pa 2610 1682}%
\special{pa 2616 1678}%
\special{pa 2620 1674}%
\special{pa 2626 1670}%
\special{pa 2630 1666}%
\special{pa 2636 1662}%
\special{pa 2640 1658}%
\special{pa 2646 1654}%
\special{pa 2650 1648}%
\special{pa 2656 1644}%
\special{pa 2660 1640}%
\special{pa 2666 1634}%
\special{pa 2670 1628}%
\special{pa 2676 1624}%
\special{pa 2680 1618}%
\special{pa 2686 1612}%
\special{pa 2690 1608}%
\special{pa 2696 1602}%
\special{pa 2700 1596}%
\special{pa 2706 1590}%
\special{pa 2710 1582}%
\special{pa 2716 1576}%
\special{pa 2720 1570}%
\special{pa 2726 1562}%
\special{pa 2730 1556}%
\special{pa 2736 1548}%
\special{pa 2740 1542}%
\special{pa 2746 1534}%
\special{pa 2750 1526}%
\special{pa 2756 1518}%
\special{pa 2760 1510}%
\special{pa 2766 1502}%
\special{pa 2770 1494}%
\special{pa 2776 1484}%
\special{pa 2780 1476}%
\special{pa 2786 1466}%
\special{pa 2790 1458}%
\special{pa 2796 1448}%
\special{pa 2800 1438}%
\special{pa 2806 1428}%
\special{pa 2810 1418}%
\special{pa 2816 1406}%
\special{pa 2820 1396}%
\special{pa 2826 1384}%
\special{pa 2830 1372}%
\special{pa 2836 1362}%
\special{pa 2840 1350}%
\special{pa 2846 1336}%
\special{pa 2850 1324}%
\special{pa 2856 1310}%
\special{pa 2860 1298}%
\special{pa 2866 1284}%
\special{pa 2870 1270}%
\special{pa 2876 1256}%
\special{pa 2880 1240}%
\special{pa 2886 1226}%
\special{pa 2890 1210}%
\special{pa 2896 1194}%
\special{pa 2900 1178}%
\special{pa 2906 1160}%
\special{pa 2910 1144}%
\special{pa 2916 1126}%
\special{pa 2920 1108}%
\special{pa 2926 1090}%
\special{pa 2930 1070}%
\special{pa 2936 1052}%
\special{pa 2940 1032}%
\special{pa 2946 1010}%
\special{pa 2950 990}%
\special{pa 2956 968}%
\special{pa 2960 946}%
\special{pa 2966 924}%
\special{pa 2970 900}%
\special{pa 2976 876}%
\special{pa 2980 852}%
\special{pa 2986 828}%
\special{pa 2990 802}%
\special{pa 2996 776}%
\special{pa 3000 748}%
\special{pa 3006 720}%
\special{pa 3010 692}%
\special{pa 3016 664}%
\special{pa 3020 634}%
\special{pa 3026 602}%
\special{pa 3030 572}%
\special{pa 3036 540}%
\special{pa 3040 506}%
\special{pa 3046 472}%
\special{pa 3050 438}%
\special{pa 3056 402}%
\special{pa 3056 400}%
\special{sp}%
% LINE 2 0 3 0
% 2 1330 400 3050 400
% 
\special{pn 8}%
\special{pa 1330 400}%
\special{pa 3050 400}%
\special{fp}%
% VECTOR 1 0 3 0
% 2 2580 2000 2270 2000
% 
\special{pn 13}%
\special{pa 2580 2000}%
\special{pa 2270 2000}%
\special{fp}%
\special{sh 1}%
\special{pa 2270 2000}%
\special{pa 2338 2020}%
\special{pa 2324 2000}%
\special{pa 2338 1980}%
\special{pa 2270 2000}%
\special{fp}%
% STR 2 0 3 0
% 3 2630 1960 2630 2060 2 0
% edge
\put(26.3000,-20.6000){\makebox(0,0)[lb]{edge}}%
% NONE 1 0 3 0
% 2 1200 1800 2000 1400
% 
% STR 2 0 3 0
% 3 1070 1880 1070 1980 2 0
% body
\put(10.7000,-19.8000){\makebox(0,0)[lb]{body}}%
% VECTOR 1 0 3 0
% 2 1280 1770 1980 1650
% 
\special{pn 13}%
\special{pa 1280 1770}%
\special{pa 1980 1650}%
\special{fp}%
\special{sh 1}%
\special{pa 1980 1650}%
\special{pa 1912 1642}%
\special{pa 1928 1660}%
\special{pa 1918 1682}%
\special{pa 1980 1650}%
\special{fp}%
% VECTOR 2 0 3 0
% 2 800 2200 3600 2200
% 
\special{pn 8}%
\special{pa 800 2200}%
\special{pa 3600 2200}%
\special{fp}%
\special{sh 1}%
\special{pa 3600 2200}%
\special{pa 3534 2180}%
\special{pa 3548 2200}%
\special{pa 3534 2220}%
\special{pa 3600 2200}%
\special{fp}%
% STR 2 0 3 0
% 3 3190 2320 3190 2420 2 0
% Re($2 \beta h$)
\put(31.9000,-24.2000){\makebox(0,0)[lb]{Re($2 \beta h$)}}%
% VECTOR 2 0 3 0
% 2 800 2200 800 200
% 
\special{pn 8}%
\special{pa 800 2200}%
\special{pa 800 200}%
\special{fp}%
\special{sh 1}%
\special{pa 800 200}%
\special{pa 780 268}%
\special{pa 800 254}%
\special{pa 820 268}%
\special{pa 800 200}%
\special{fp}%
% LINE 0 0 3 0
% 2 2200 1800 2200 2000
% 
\special{pn 20}%
\special{pa 2200 1800}%
\special{pa 2200 2000}%
\special{fp}%
% LINE 3 0 3 0
% 40 2680 400 1620 1460 2800 400 1670 1530 2920 400 1720 1600 3040 400 1780 1660 3020 540 1850 1710 3000 680 1930 1750 2970 830 2030 1770 2940 980 2130 1790 2900 1140 2250 1790 2830 1330 2390 1770 2560 400 1580 1380 2440 400 1550 1290 2320 400 1520 1200 2200 400 1490 1110 2080 400 1470 1010 1960 400 1440 920 1840 400 1420 820 1720 400 1410 710 1600 400 1390 610 1480 400 1370 510
% 
\special{pn 4}%
\special{pa 2680 400}%
\special{pa 1620 1460}%
\special{fp}%
\special{pa 2800 400}%
\special{pa 1670 1530}%
\special{fp}%
\special{pa 2920 400}%
\special{pa 1720 1600}%
\special{fp}%
\special{pa 3040 400}%
\special{pa 1780 1660}%
\special{fp}%
\special{pa 3020 540}%
\special{pa 1850 1710}%
\special{fp}%
\special{pa 3000 680}%
\special{pa 1930 1750}%
\special{fp}%
\special{pa 2970 830}%
\special{pa 2030 1770}%
\special{fp}%
\special{pa 2940 980}%
\special{pa 2130 1790}%
\special{fp}%
\special{pa 2900 1140}%
\special{pa 2250 1790}%
\special{fp}%
\special{pa 2830 1330}%
\special{pa 2390 1770}%
\special{fp}%
\special{pa 2560 400}%
\special{pa 1580 1380}%
\special{fp}%
\special{pa 2440 400}%
\special{pa 1550 1290}%
\special{fp}%
\special{pa 2320 400}%
\special{pa 1520 1200}%
\special{fp}%
\special{pa 2200 400}%
\special{pa 1490 1110}%
\special{fp}%
\special{pa 2080 400}%
\special{pa 1470 1010}%
\special{fp}%
\special{pa 1960 400}%
\special{pa 1440 920}%
\special{fp}%
\special{pa 1840 400}%
\special{pa 1420 820}%
\special{fp}%
\special{pa 1720 400}%
\special{pa 1410 710}%
\special{fp}%
\special{pa 1600 400}%
\special{pa 1390 610}%
\special{fp}%
\special{pa 1480 400}%
\special{pa 1370 510}%
\special{fp}%
% STR 2 0 3 0
% 3 510 70 510 170 2 0
% Im($2 \beta h$)
\put(5.1000,-1.7000){\makebox(0,0)[lb]{Im($2 \beta h$)}}%
% STR 2 0 3 0
% 3 580 340 580 440 2 0
% $\pi i$
\put(5.8000,-4.4000){\makebox(0,0)[lb]{$\pi i$}}%
% LINE 2 0 3 0
% 2 2200 2200 2200 2100
% 
\special{pn 8}%
\special{pa 2200 2200}%
\special{pa 2200 2100}%
\special{fp}%
% LINE 2 0 3 0
% 2 800 400 900 400
% 
\special{pn 8}%
\special{pa 800 400}%
\special{pa 900 400}%
\special{fp}%
% STR 2 0 3 0
% 3 660 2070 660 2170 5 0
% $0$
\put(6.6000,-21.7000){\makebox(0,0){$0$}}%
\end{picture}%

\caption{Schematic diagram for the distribution of zeros on the complex field plane.}\label{fig0}
\end{center}
\end{minipage}
\end{figure}

The free energy for a system with quenched randomness is defined as the average over samples for all possible bond configurations. This implies that the partition function of a quenched system is regarded as being given by the product of partition functions of all possible configurations,
\begin{equation}
 - \beta F = 2^{ - N_b } \left\{ \log \prod\limits_{\mbox{\boldmath $J$}} {Z\left( {\mbox{\boldmath $J$}} \right)} \right\},
\end{equation}
where $F$ denotes the quenched free energy and $\mbox{\boldmath $J$}$ denotes the set of random interactions. This fact allows us to plot zeros of all randomly-chosen samples simultaneously on the complex plane of $\log y = 2\beta h$. The temperature will be measured in units of $J/k_{B}$. A schematic diagram for the distribution of zeros is presented in figure \ref{fig0}. 

\begin{figure}[tbp]
\begin{minipage}[t]{.5\textwidth}
\begin{center}
\includegraphics[width=1.0\hsize]{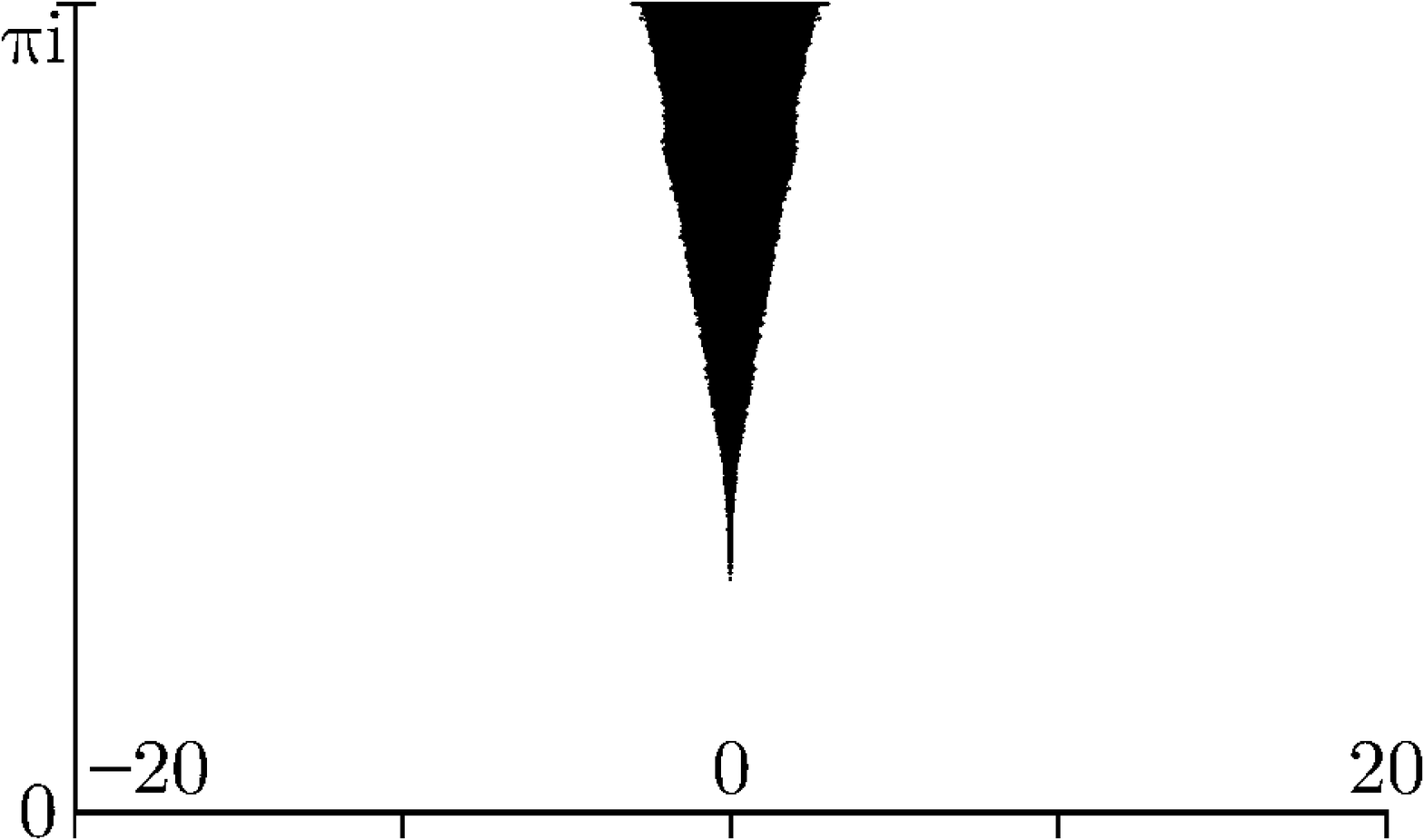}
{\footnotesize  $T=5.0$, $S=26430$}
\end{center}
\end{minipage}
\begin{minipage}[t]{.5\textwidth}
\begin{center}
\includegraphics[width=1.0\hsize]{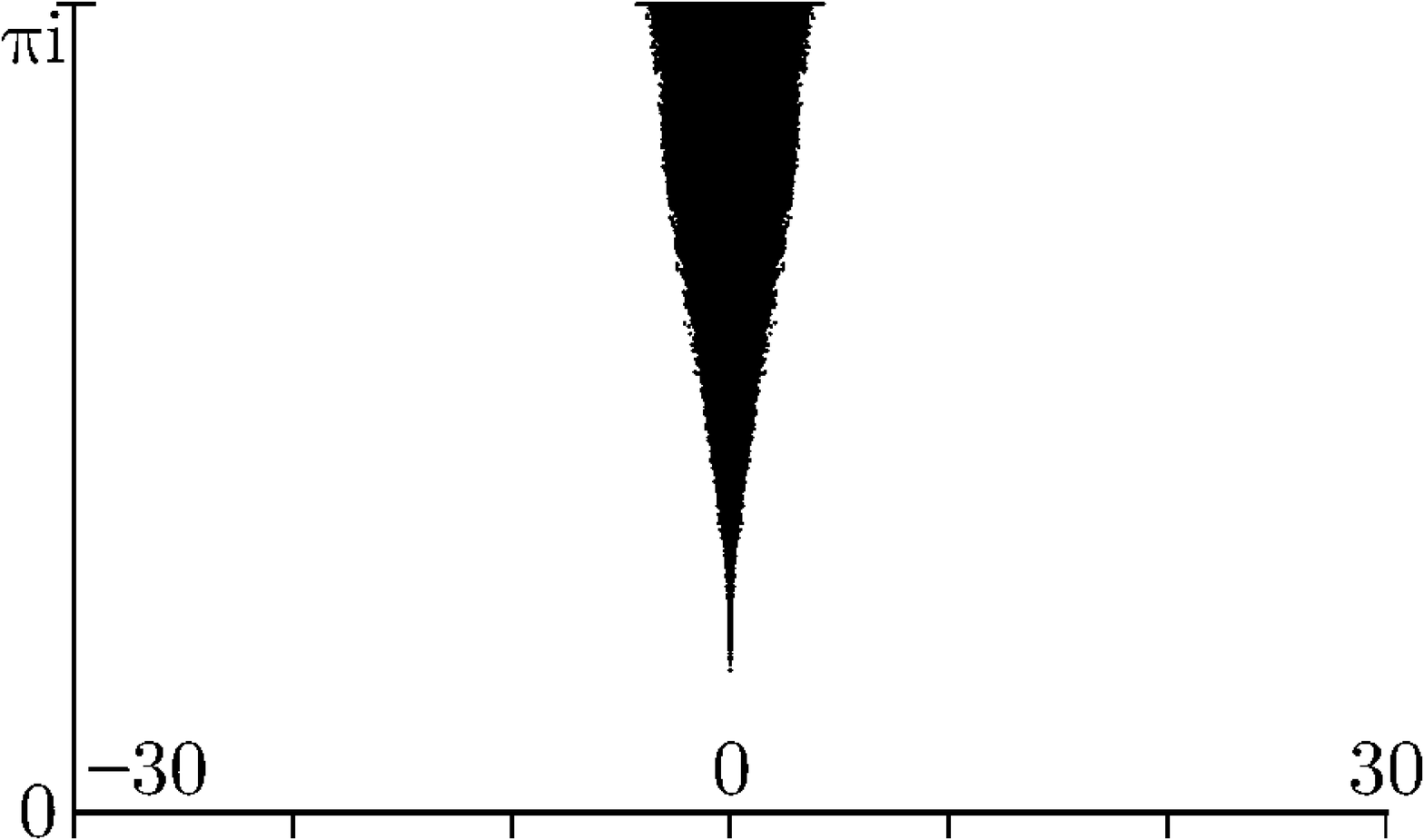}
{\footnotesize  $T=4.0$, $S=9662$}
\end{center}
\end{minipage}
\\
\\
\\
\begin{minipage}[t]{.5\textwidth}
\begin{center}
\includegraphics[width=1.0\hsize]{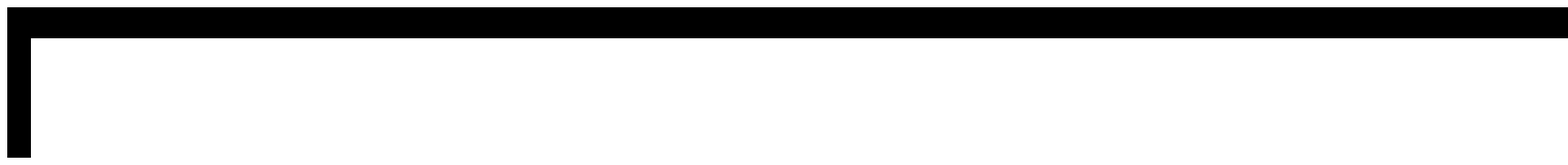}
{\footnotesize  $T=2.0$, $S=20284$}
\end{center}
\end{minipage}
\hfill
\begin{minipage}[t]{.5\textwidth}
\begin{center}
\includegraphics[width=1.0\hsize]{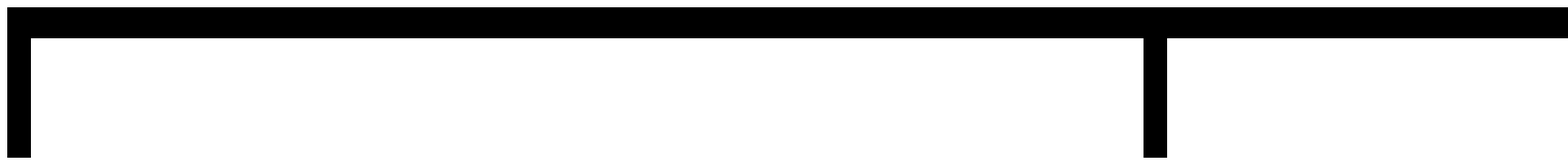}
{\footnotesize  $T=2.0$, $S=9831$}
\end{center}
\end{minipage}
\\
\\
\\
\begin{minipage}[t]{.5\textwidth}
\begin{center}
\includegraphics[width=1.0\hsize]{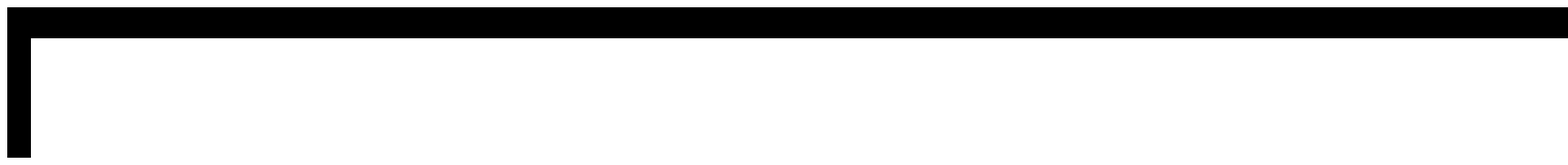}
{\footnotesize  $T=1.0$, $S=14536$}
\end{center}
\end{minipage}
\hfill
\begin{minipage}[t]{.5\textwidth}
\begin{center}
\includegraphics[width=1.0\hsize]{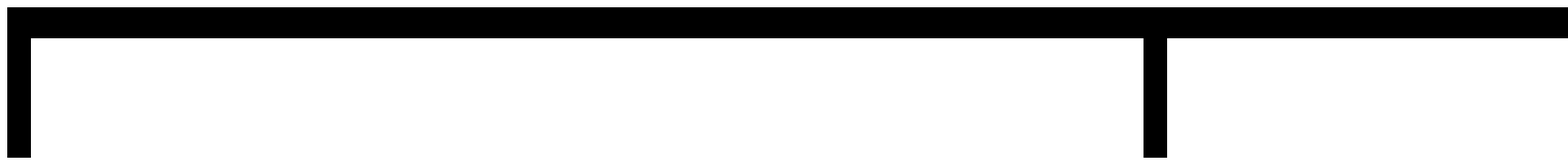}
{\footnotesize  $T=1.0$, $S=10521$}
\end{center}
\end{minipage}
\\
\\
\\
\begin{minipage}[t]{.5\textwidth}
\begin{center}
\includegraphics[width=1.0\hsize]{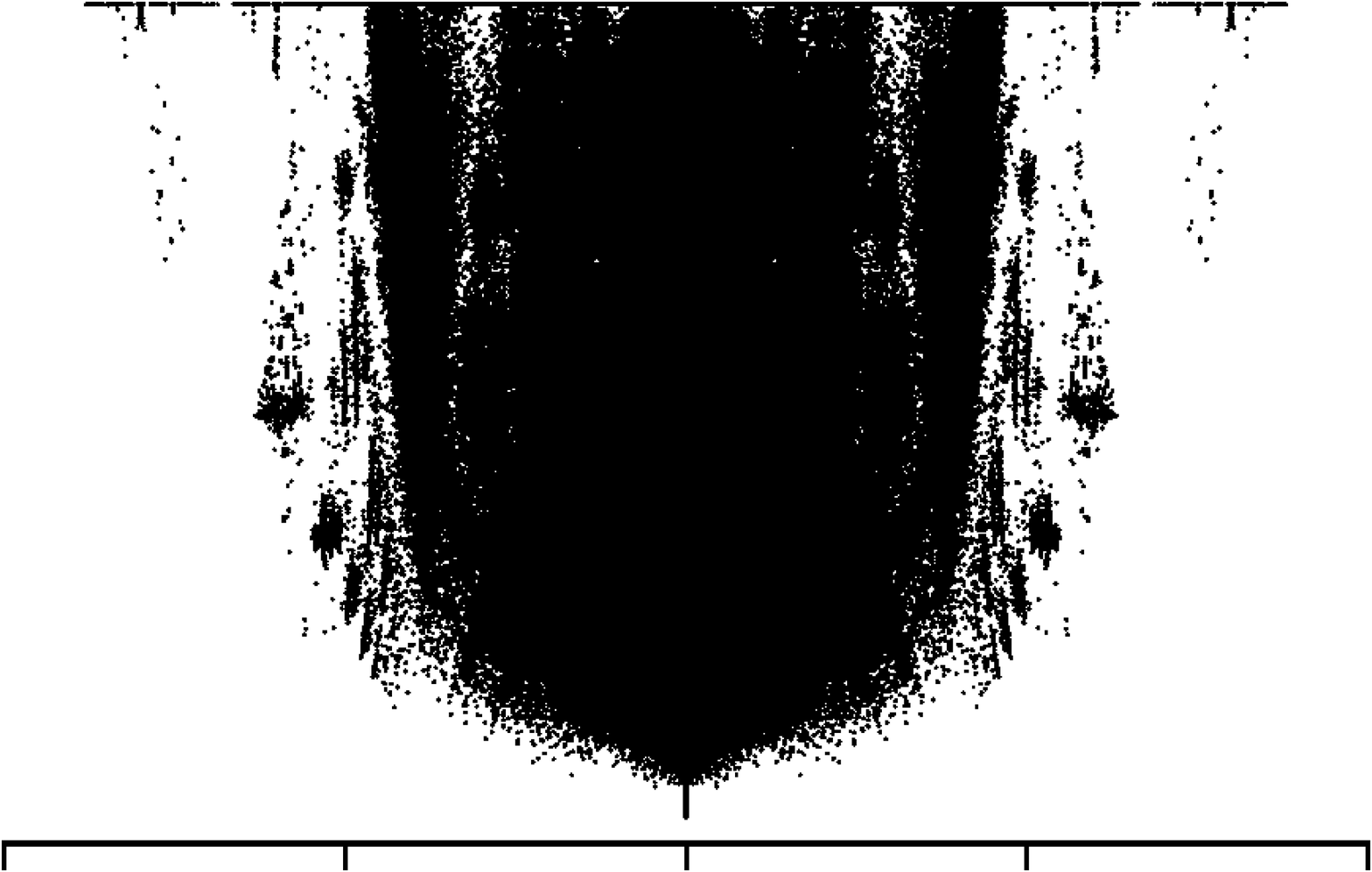}
{\footnotesize  $T=0.5$, $S=17901$}
\end{center}
\end{minipage}
\hfill
\begin{minipage}[t]{.5\textwidth}
\begin{center}
\includegraphics[width=1.0\hsize]{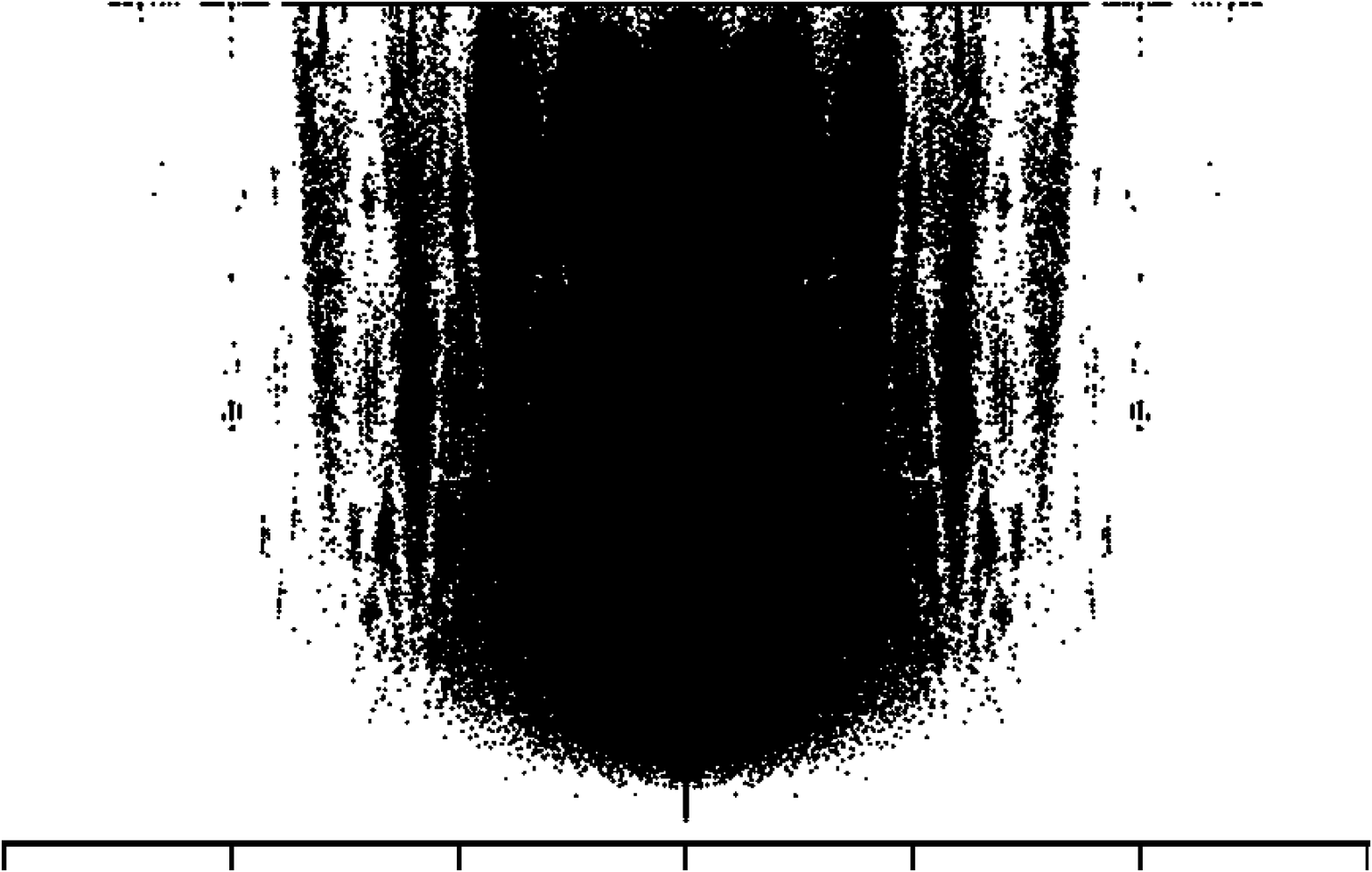}
{\footnotesize  $T=0.5$, $S=12083$}
\end{center}
\end{minipage}
\begin{center}
\caption{Temperature dependence of the distribution of zeros on the complex $2\beta h$ plane. The system sizes are $10 \times 10$ (left figures) and $4\times4\times6$ (right figures). $S$ denotes the number of samples.}\label{fig1}
\end{center}
\end{figure}

Figure \ref{fig1} shows the temperature dependence of the distribution of zeros for systems of size $10 \times 10$ in two dimensions and $4 \times 4 \times 6$ in three dimensions. Throughout this paper the range of the real axis is from $-20$ to $20$ in two dimensions and from $-30$ to $30$ in three dimensions, and the imaginary axis ranges from $0$ to $\pi i$.

Generally, zeros of both dimensions approach the real axis with decreasing temperature. Nearest zeros to the origin lie on the imaginary axis for both dimensions and all temperatures. Let us call the location of such zeros the \textit{edge} (see figure \ref{fig0}). A clear difference between two and three dimensions appears at low temperature. Zeros lying off the imaginary axis but close to the origin (to be called the \textit{body}) form a wedged-like shape in two dimensions whereas the body looks relatively rounded in three dimensions.

\begin{figure}[tbp]
\begin{minipage}[t]{.5\textwidth}
\begin{center}
\includegraphics[width=1.0\hsize]{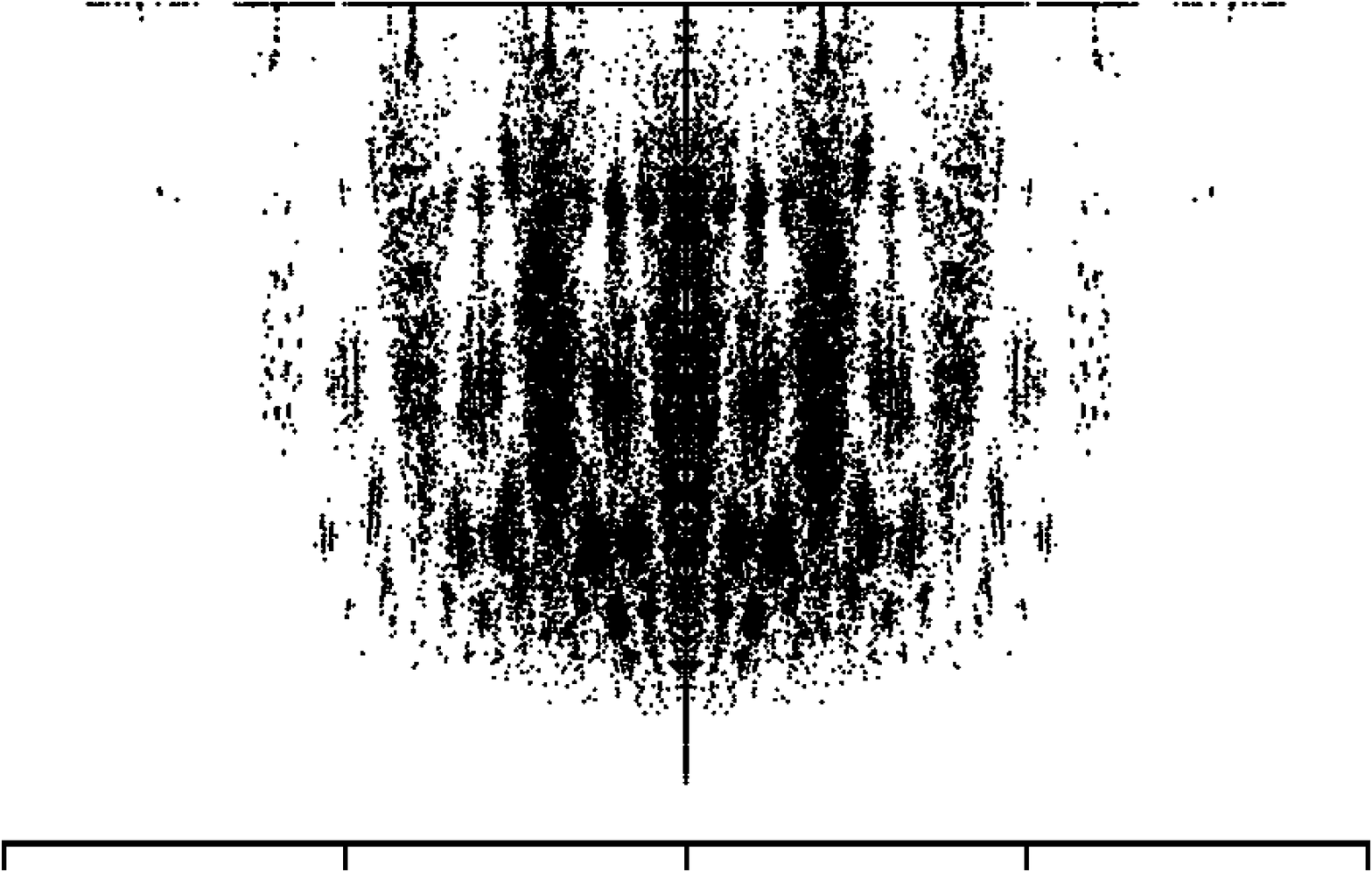}
{\footnotesize  $4 \times 4$, $S=20496$}
\end{center}
\end{minipage}
\begin{minipage}[t]{.5\textwidth}
\begin{center}
\includegraphics[width=1.0\hsize]{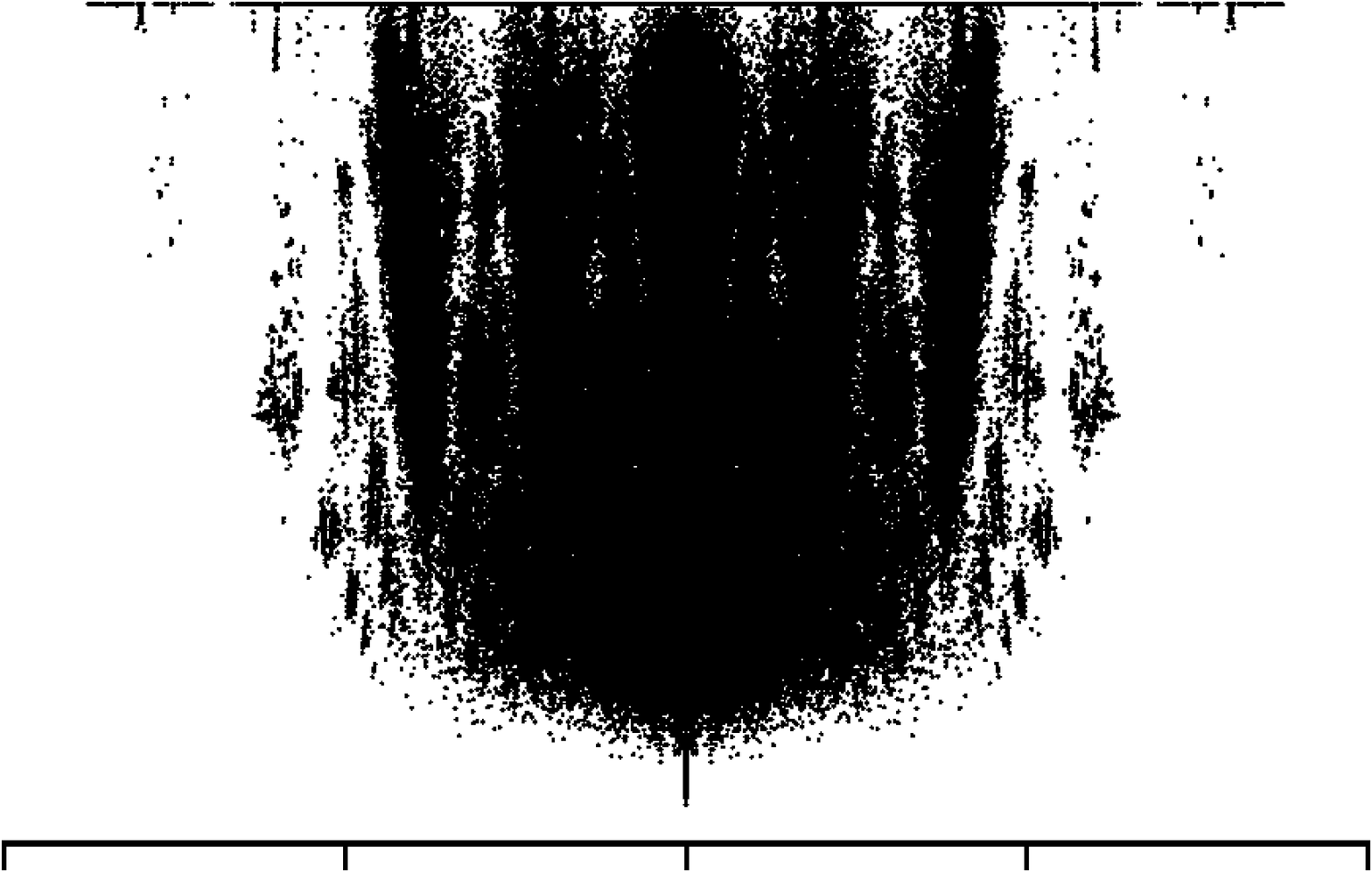}
{\footnotesize  $6 \times 6$, $S=34943$}
\end{center}
\end{minipage}
\\
\\
\\
\begin{minipage}[t]{.5\textwidth}
\begin{center}
\includegraphics[width=1.0\hsize]{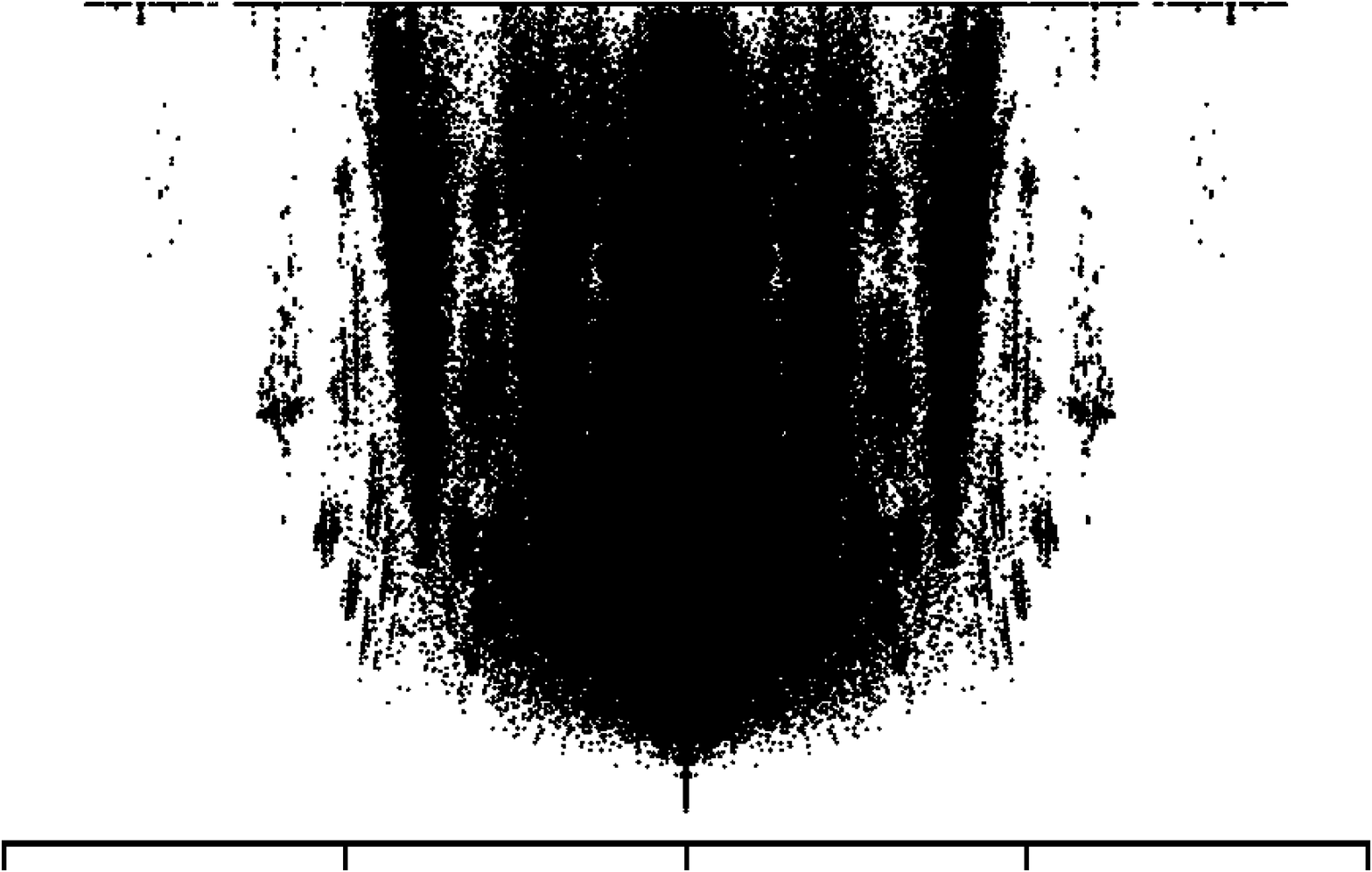}
{\footnotesize  $8 \times 8$, $S=14290$}
\end{center}
\end{minipage}
\hfill
\begin{minipage}[t]{.5\textwidth}
\begin{center}
\includegraphics[width=1.0\hsize]{fig1_2_10_05_17901.eps}
{\footnotesize  $10 \times 10$, $S=17901$}
\end{center}
\end{minipage}
\\
\\
\\
\begin{minipage}[t]{.5\textwidth}
\begin{center}
\includegraphics[width=1.0\hsize]{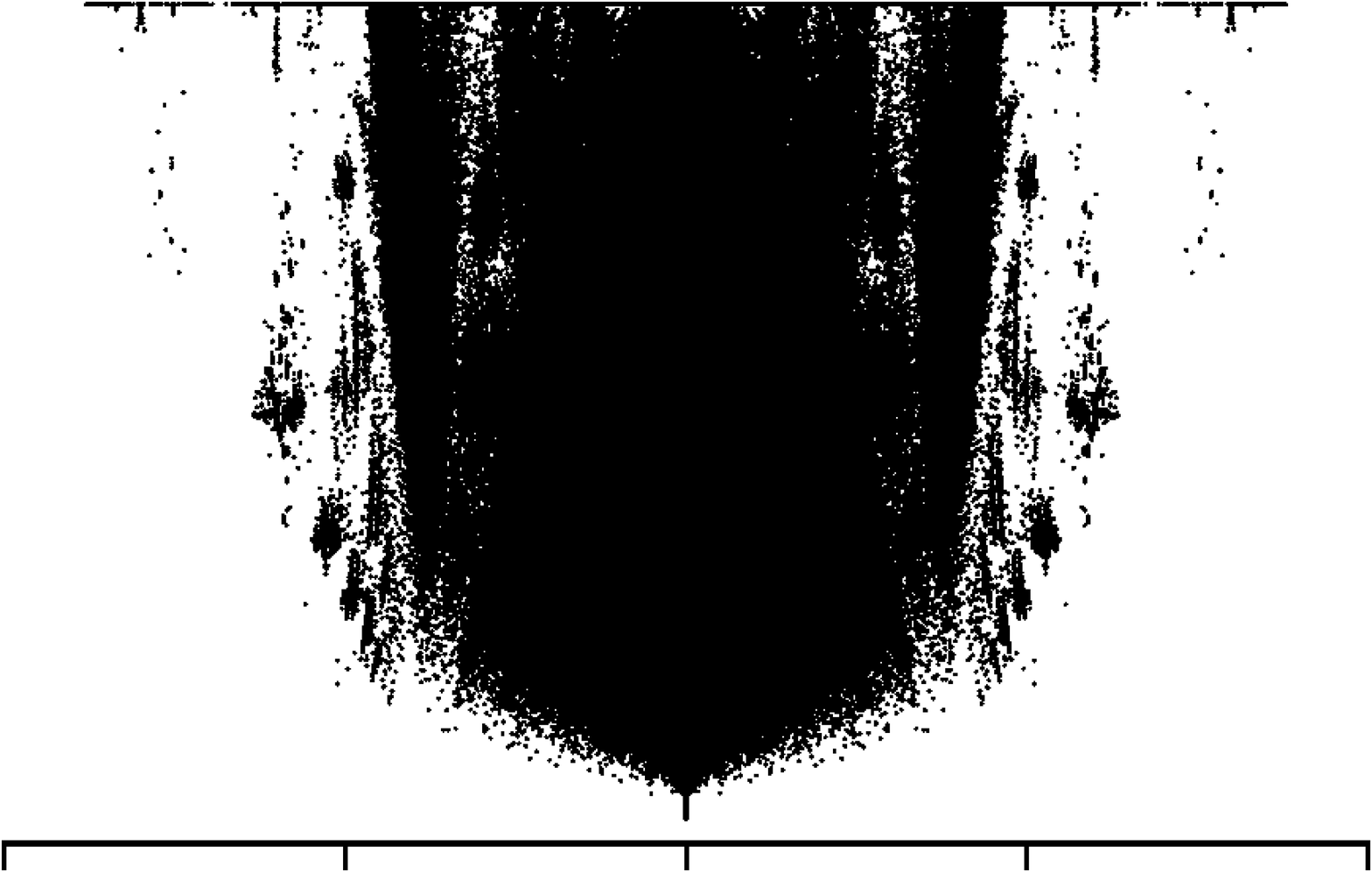}
{\footnotesize  $12 \times 12$, $S=12251$}
\end{center}
\end{minipage}
\hfill
\begin{minipage}[t]{.5\textwidth}
\begin{center}
\includegraphics[width=1.0\hsize]{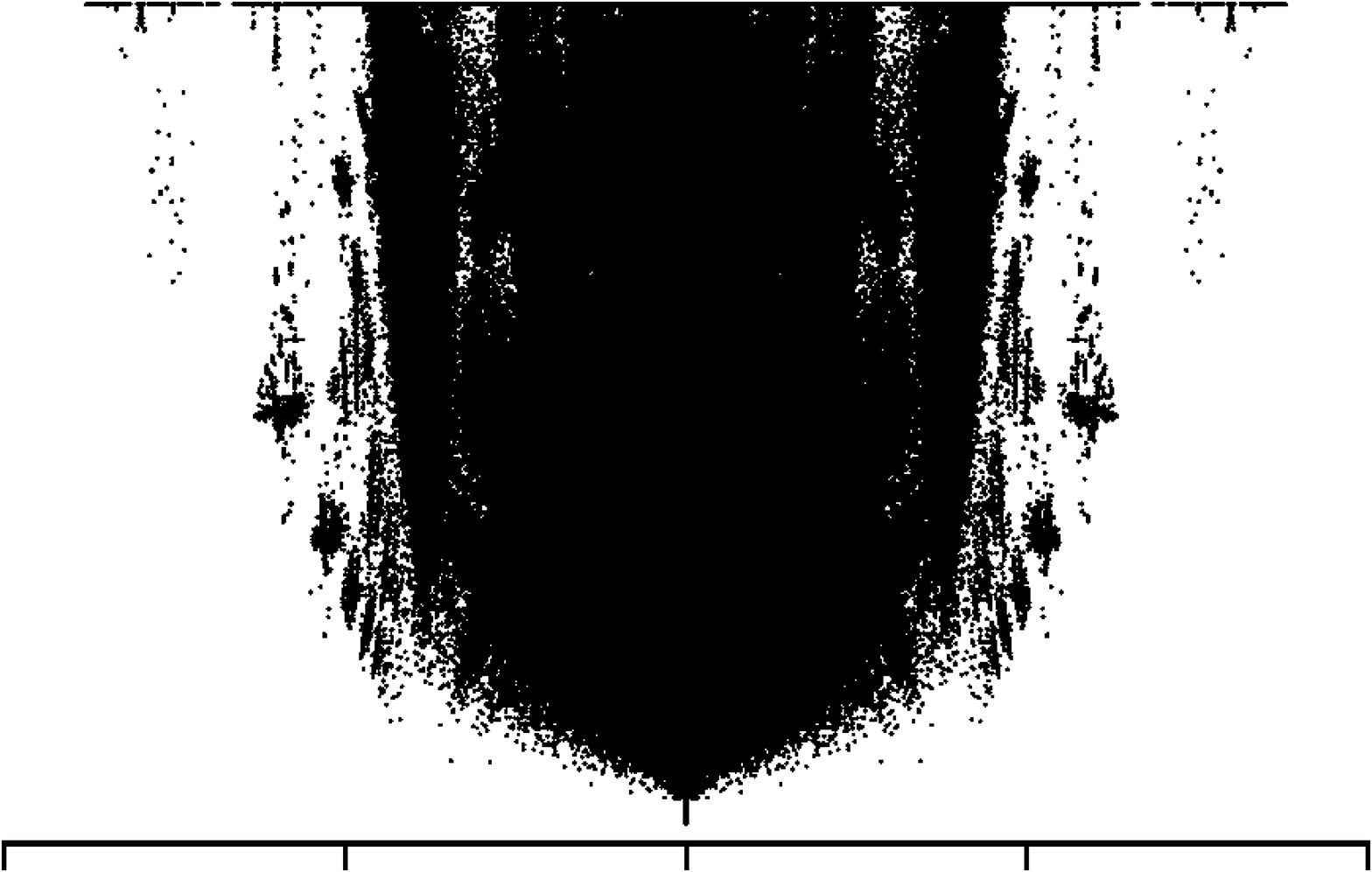}
{\footnotesize  $14 \times 14$, $S=11148$}
\end{center}
\end{minipage}
\\
\\
\\
\begin{minipage}[t]{.5\textwidth}
\begin{center}
\includegraphics[width=1.0\hsize]{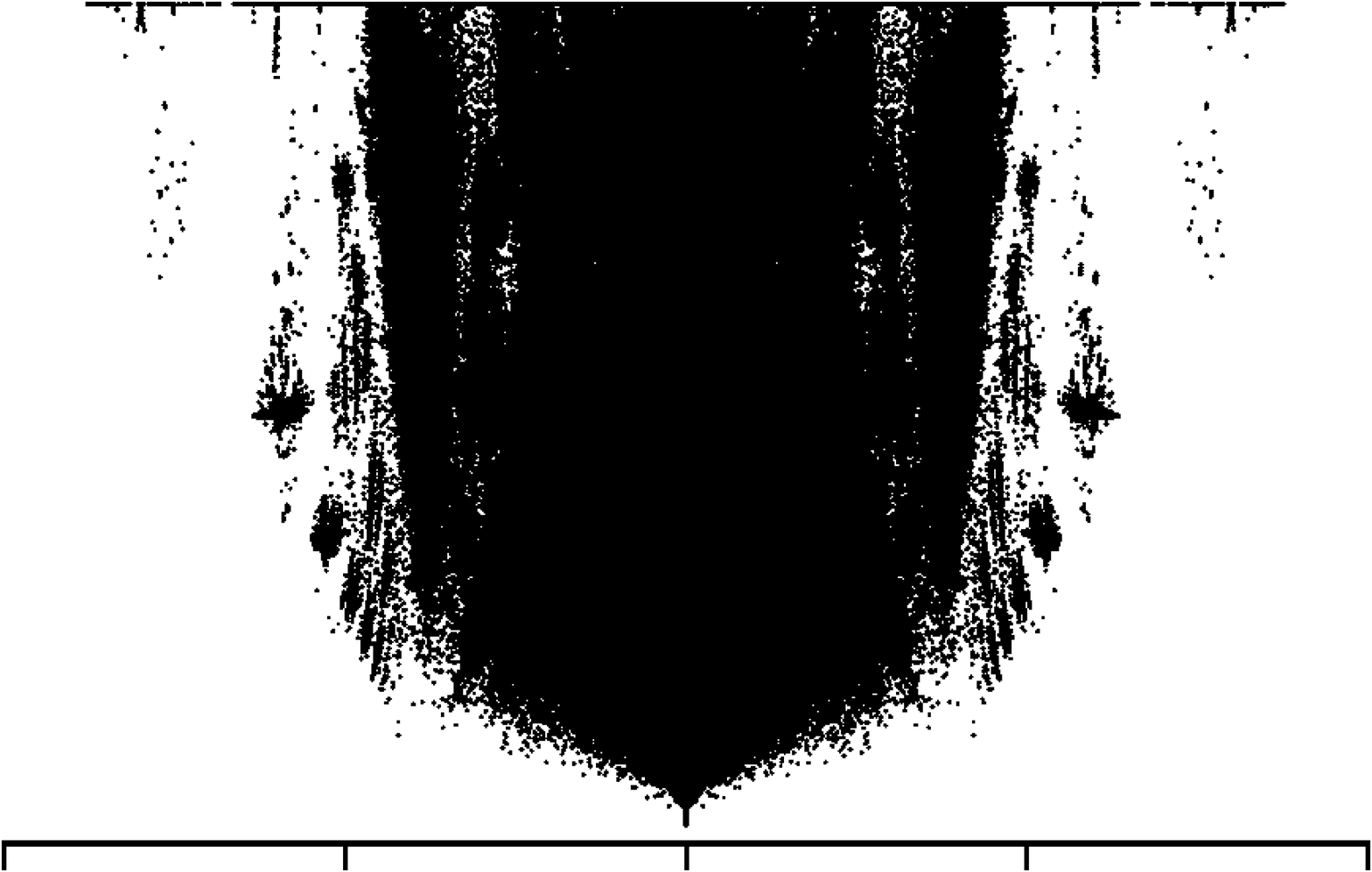}
{\footnotesize  $16 \times 16$, $S=9258$}
\end{center}
\end{minipage}
\hfill
\begin{minipage}[t]{.5\textwidth}
\end{minipage}
\begin{center}
\caption{System-size dependence in two dimensions at $T=0.5$.}\label{fig2}
\end{center}
\end{figure}

\begin{figure}[tbp]
\begin{minipage}[t]{.5\textwidth}
\begin{center}
\includegraphics[width=1.0\hsize]{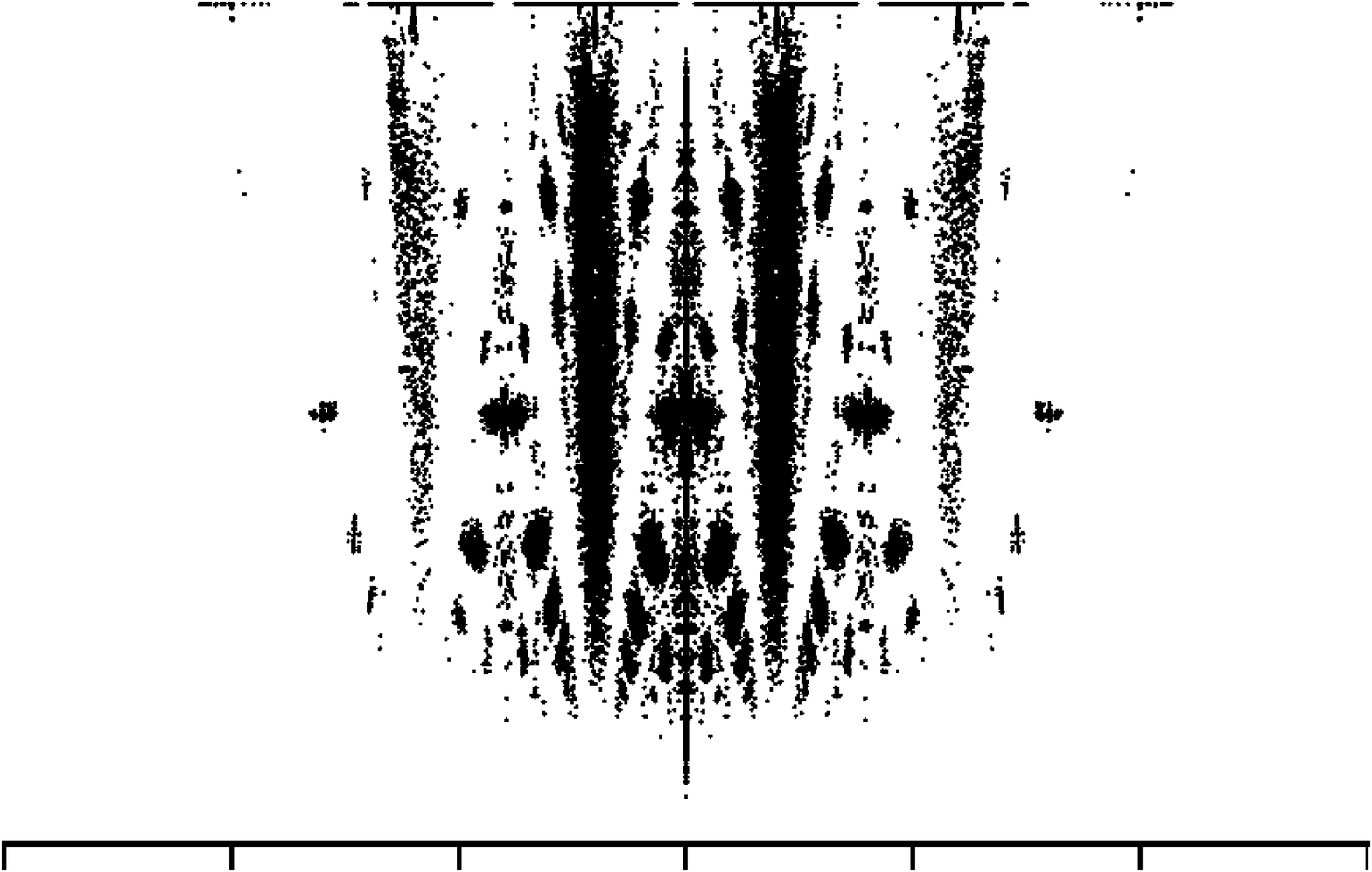}
{\footnotesize  $3\times3\times2$, $S=20892$}
\end{center}
\end{minipage}
\begin{minipage}[t]{.5\textwidth}
\begin{center}
\includegraphics[width=1.0\hsize]{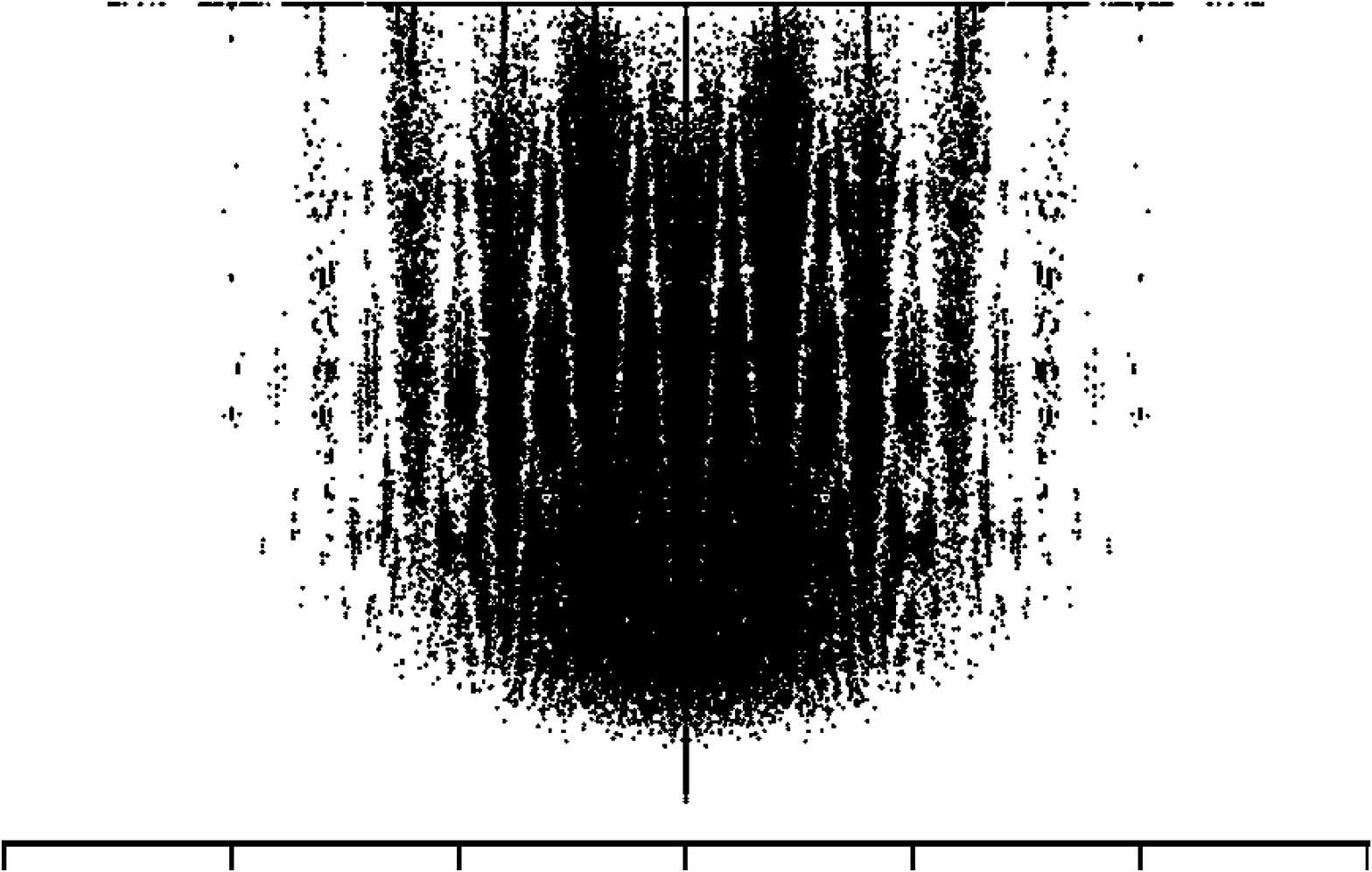}
{\footnotesize  $3\times3\times3$, $S=23829$}
\end{center}
\end{minipage}
\\
\\
\\
\begin{minipage}[t]{.5\textwidth}
\begin{center}
\includegraphics[width=1.0\hsize]{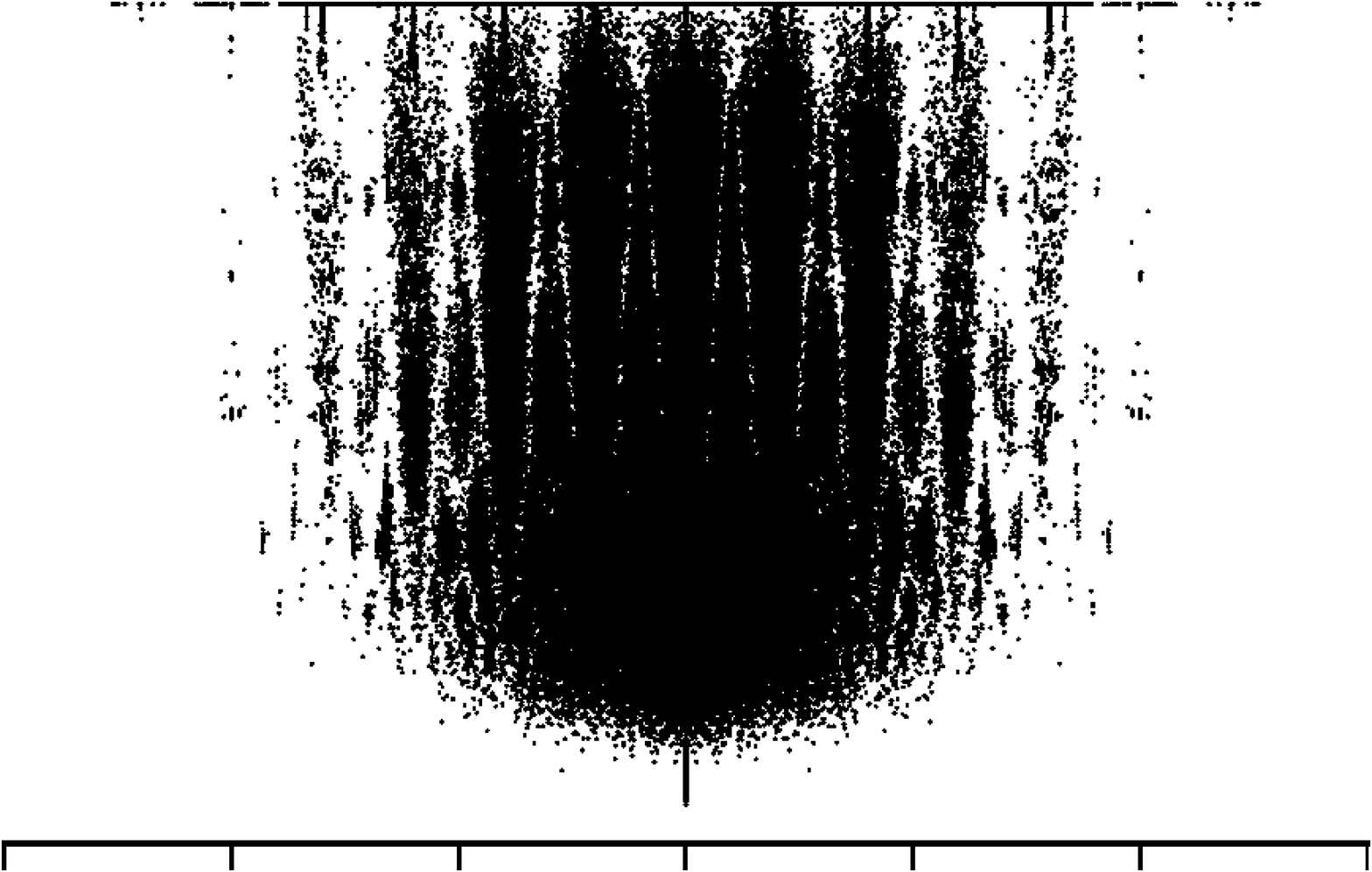}
{\footnotesize  $3\times3\times4$, $S=21510$}
\end{center}
\end{minipage}
\hfill
\begin{minipage}[t]{.5\textwidth}
\begin{center}
\includegraphics[width=1.0\hsize]{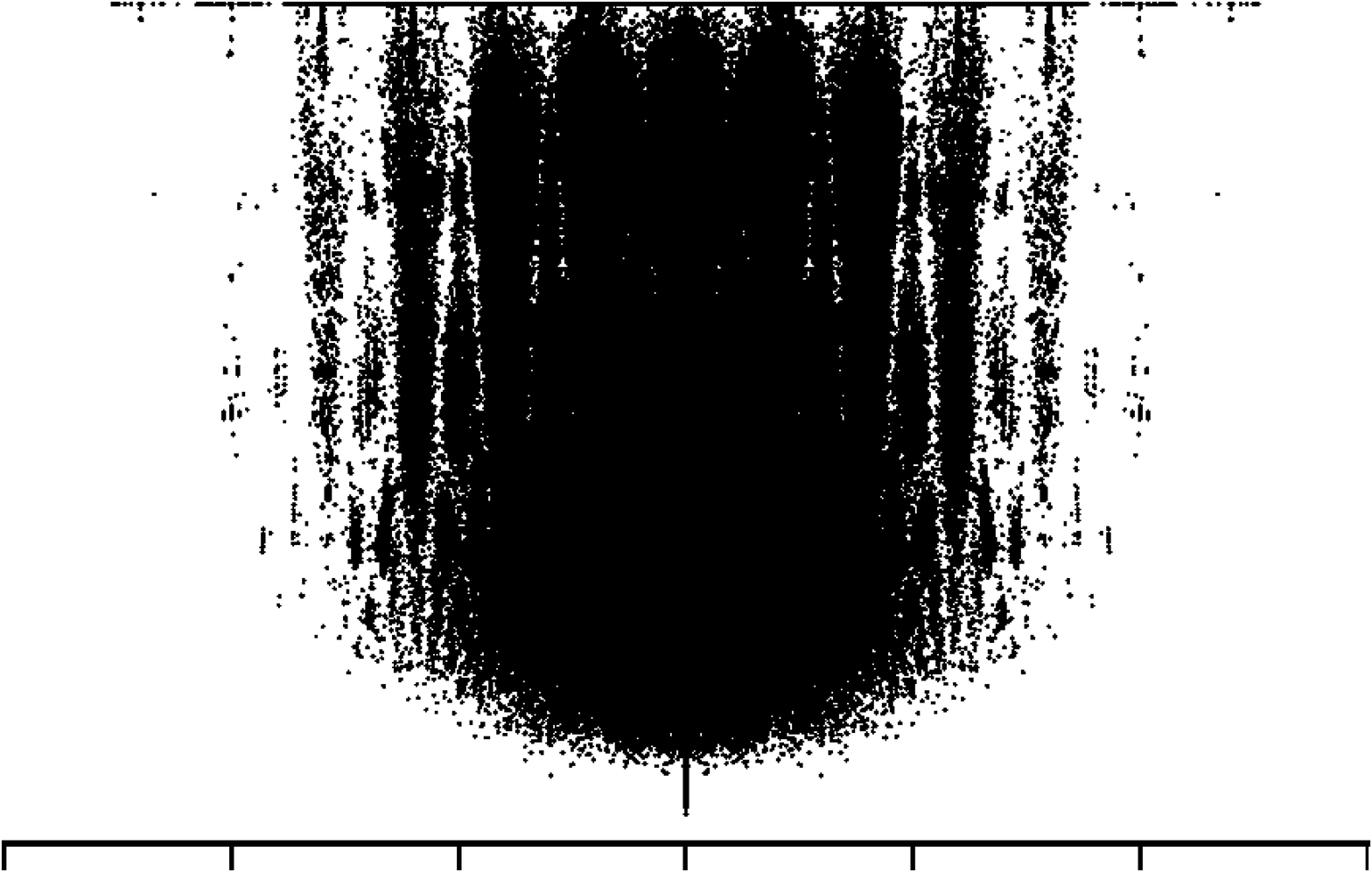}
{\footnotesize  $3\times4\times4$, $S=25808$}
\end{center}
\end{minipage}
\\
\\
\\
\begin{minipage}[t]{.5\textwidth}
\begin{center}
\includegraphics[width=1.0\hsize]{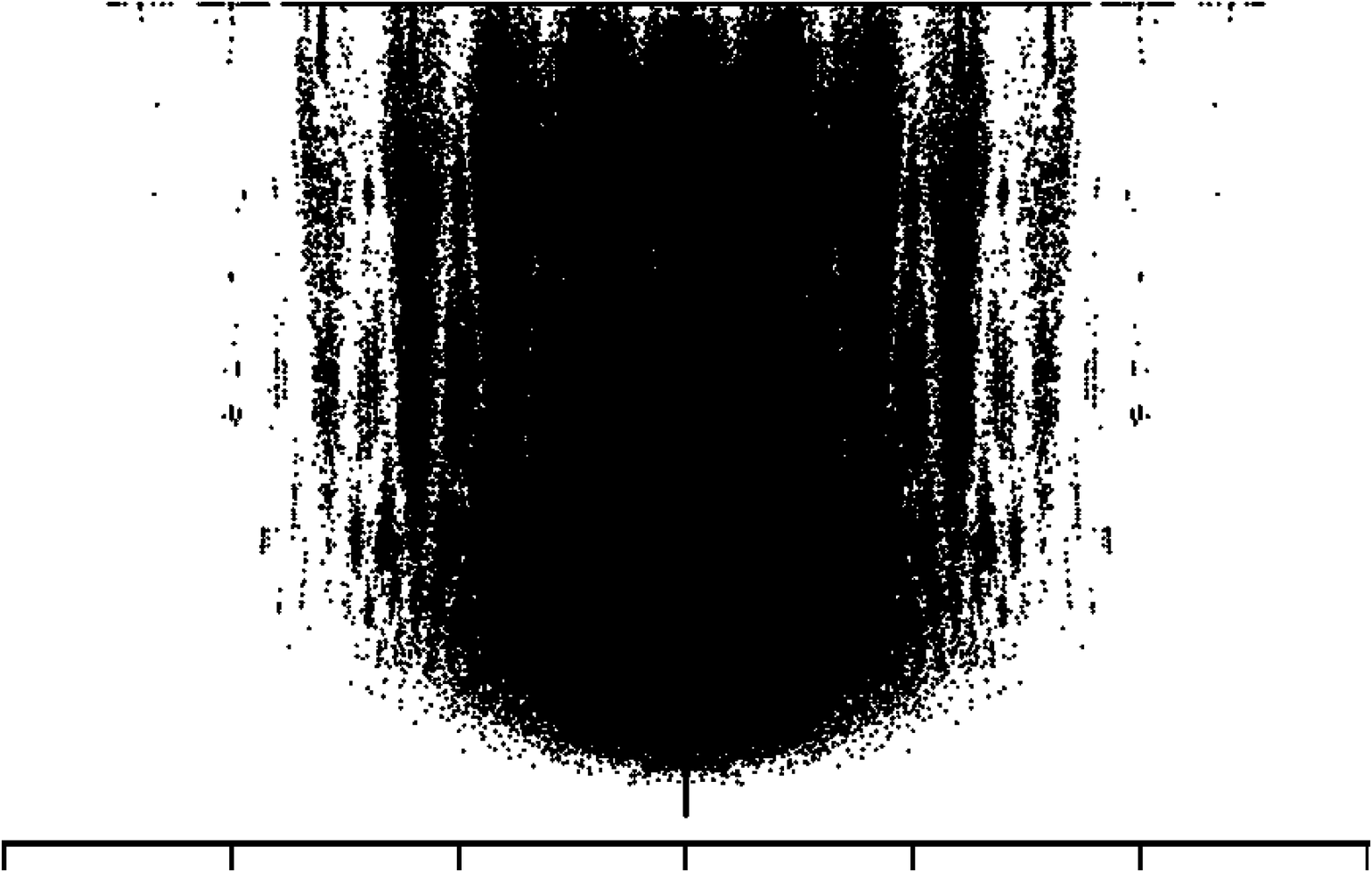}
{\footnotesize  $4\times4\times4$, $S=21501$}
\end{center}
\end{minipage}
\hfill
\begin{minipage}[t]{.5\textwidth}
\begin{center}
\includegraphics[width=1.0\hsize]{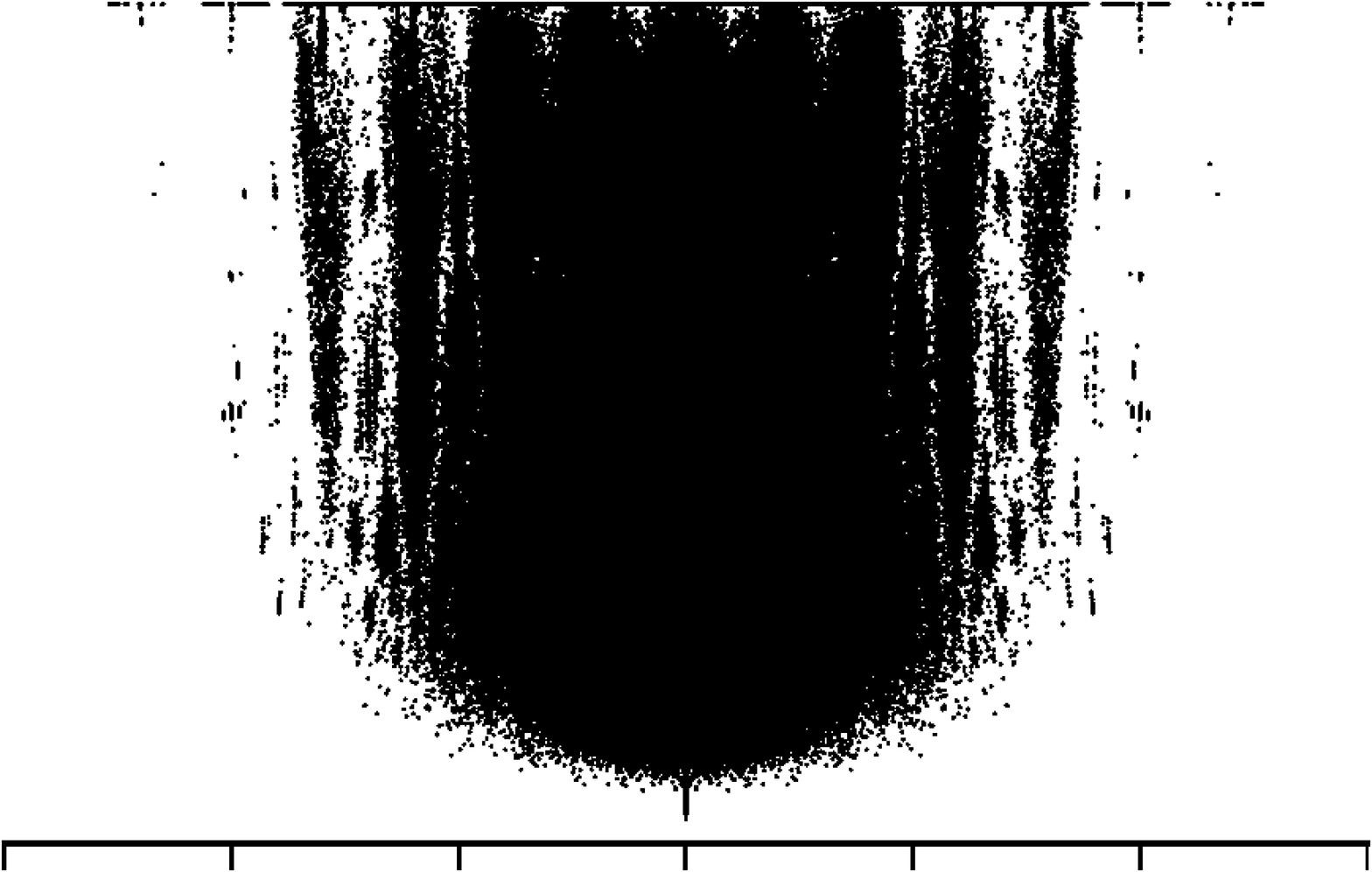}
{\footnotesize  $4\times4\times5$, $S=21568$}
\end{center}
\end{minipage}
\\
\\
\\
\begin{minipage}[t]{.5\textwidth}
\begin{center}
\includegraphics[width=1.0\hsize]{fig1_3_446_05_12083.eps}
{\footnotesize  $4\times4\times6$, $S=12083$}
\end{center}
\end{minipage}
\hfill
\begin{minipage}[t]{.5\textwidth}
\end{minipage}
\begin{center}
\caption{System-size dependence in three dimensions at $T=0.5$.}\label{fig3}
\end{center}
\end{figure}

We next discuss the dependence of the distribution on system size. Figures \ref{fig2} and \ref{fig3} show the system-size dependence at $T=0.5$ from system size $4 \times 4$ to $16 \times 16$ in two dimensions (figure \ref{fig2}) and from $3 \times 3 \times 2$ to $4 \times 4 \times 6$ in three dimensions (figure \ref{fig3}). The width of the distribution in the transverse direction does not change significantly as the system size increases and is likely to stay almost the same for sufficiently large systems. Both the edge and the body move toward the real axis as the system size increases in both dimensions. Especially, the body ({\it i.e.} zeros off the imaginary axis) tends to move more significantly than the edge. In the thermodynamic limit, if the body reaches the real axis away from the origin, it may imply that an ordered phase (that may exist along the real axis at and around the origin) would survive until the field exceeds a finite value corresponding to the point where the zero (the body) reaches the real axis.

\begin{figure}[tp]
\begin{minipage}[t]{1.0\textwidth}
  \begin{center}
   \includegraphics[width=1.0\hsize]{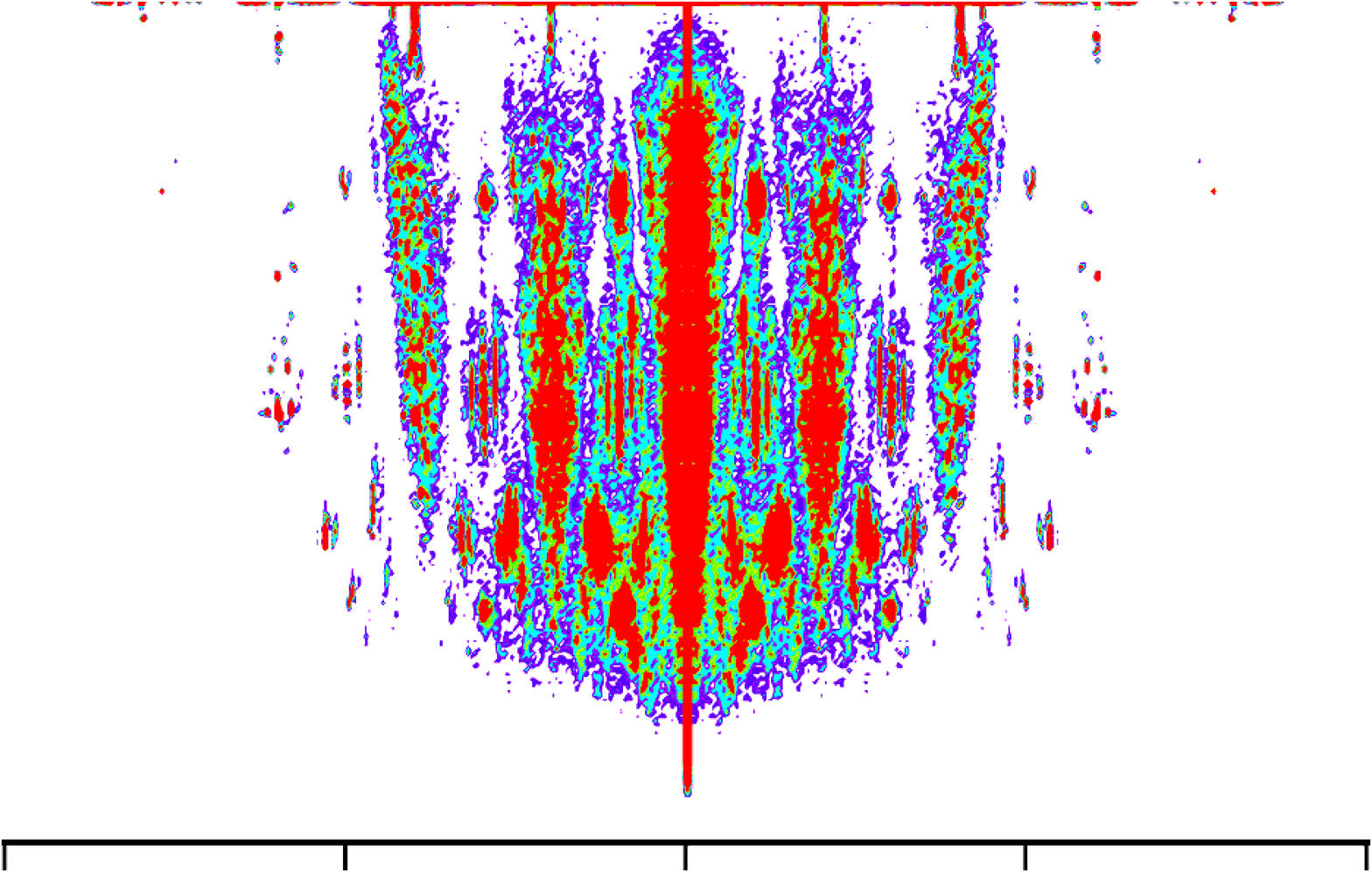}
\end{center}
\end{minipage}
\\
\\
\\
\begin{minipage}[t]{1.0\textwidth}
  \begin{center}
   \includegraphics[width=1.0\hsize]{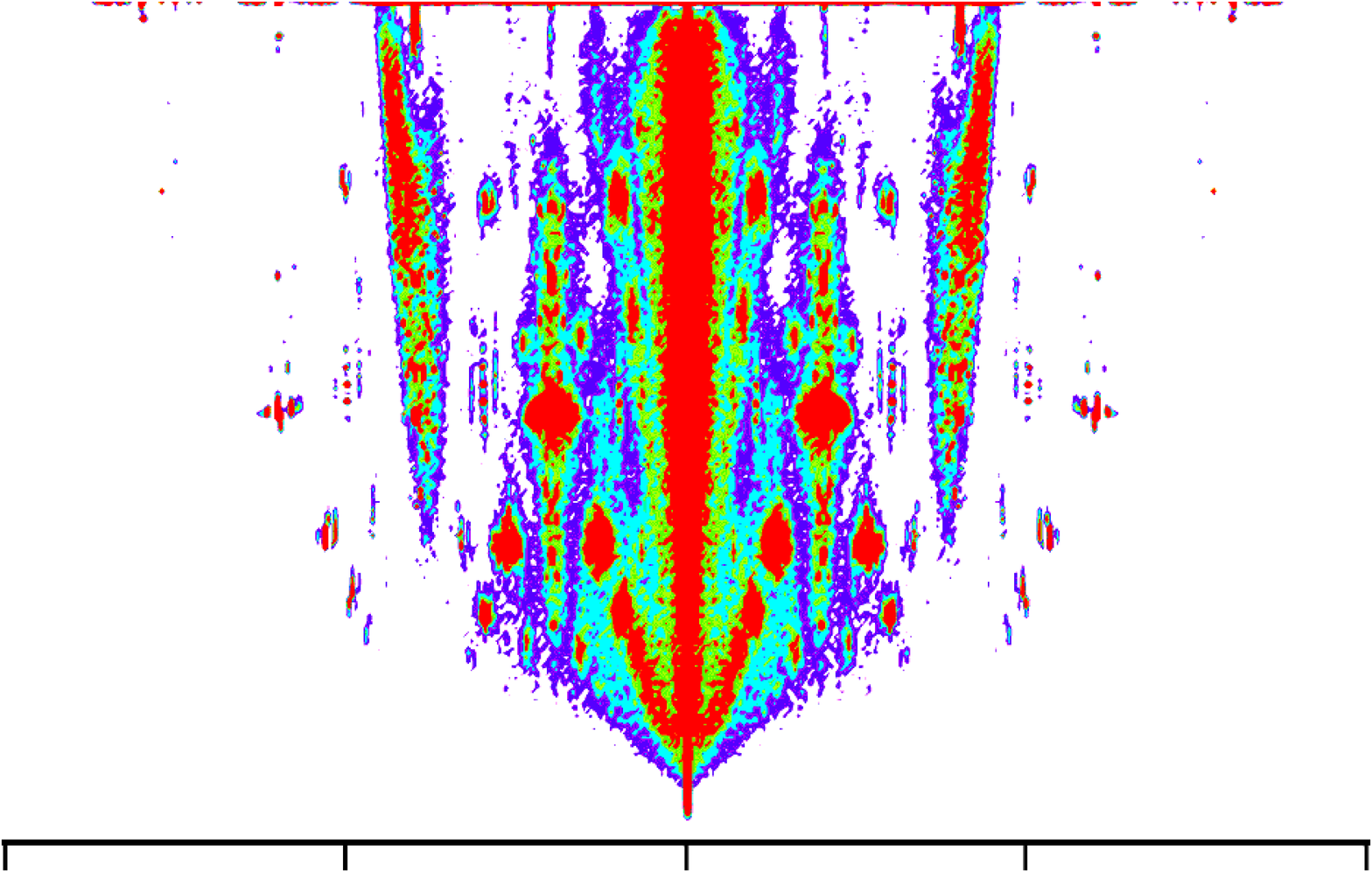}
\end{center}
\end{minipage}
\caption{Density plot of the distribution of zeros in two dimensions at $T=0.5$.  Colours represent the density: red (highest), green, light blue, purple and white (lowest) in this order. The system size is $6 \times 6$ (top) and $16 \times 16$ (bottom).}\label{fig4}
\end{figure}

\begin{figure}[tp]
\begin{minipage}[t]{1.0\textwidth}
  \begin{center}
   \includegraphics[width=1.0\hsize]{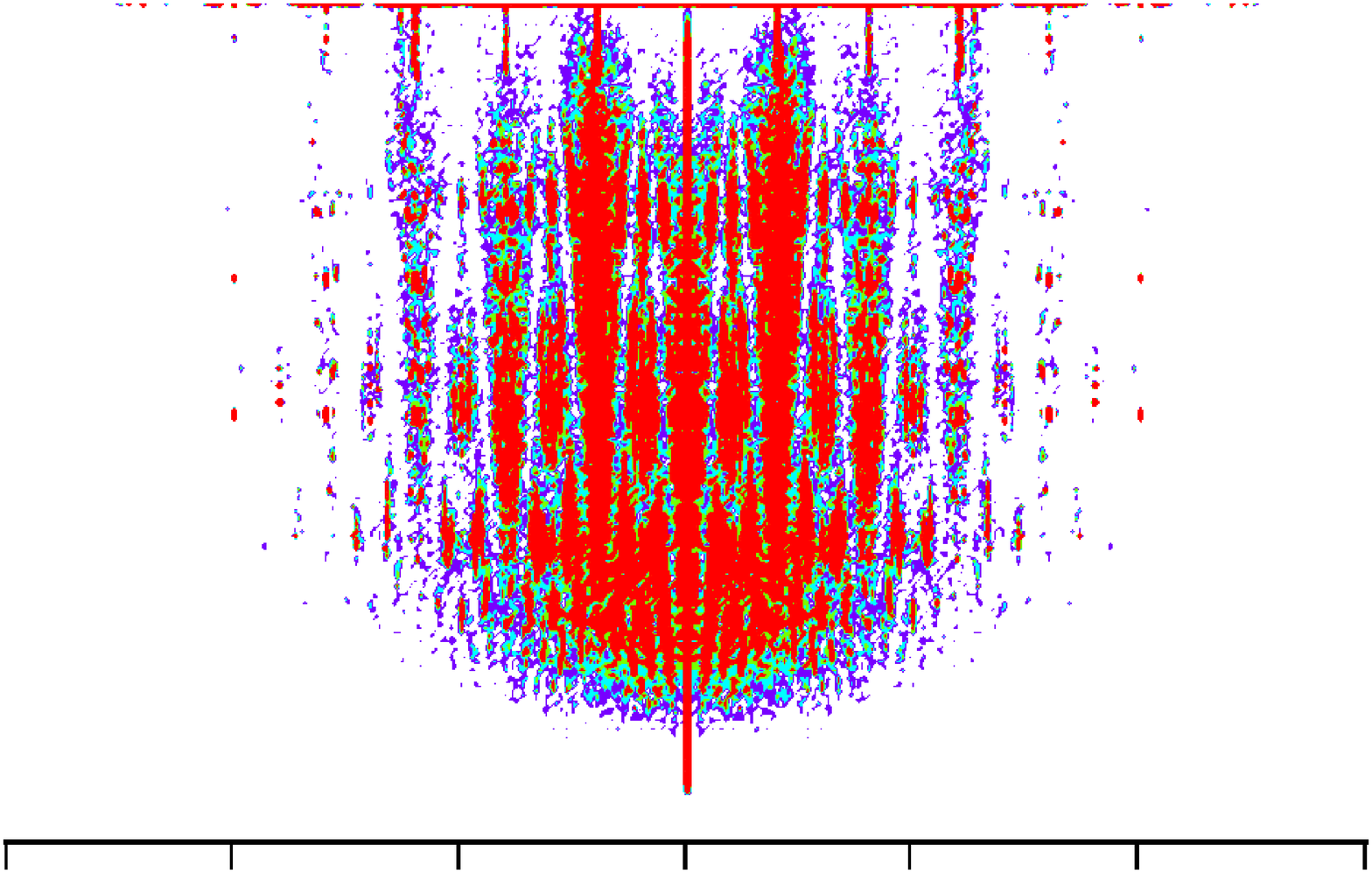}
\end{center}
\end{minipage}
\\
\\
\\
\begin{minipage}[t]{1.0\textwidth}
\begin{center}
   \includegraphics[width=1.0\hsize]{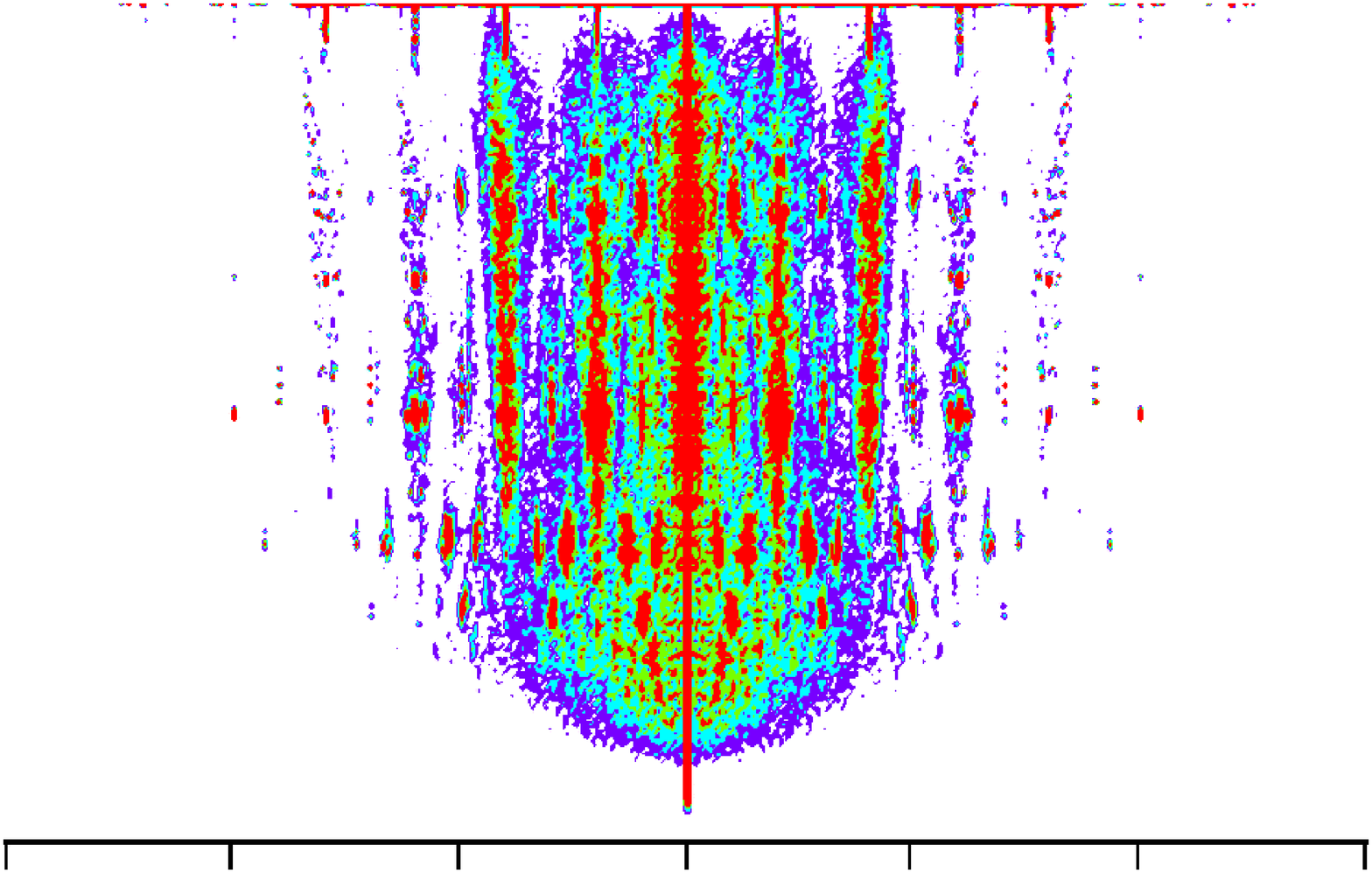}
\end{center}
\end{minipage}
\caption{Density plot of the distribution of zeros in three dimensions at $T=0.5$. The same colour code is used as in figure \ref{fig4}. The system size is $3 \times 3\times3$ (top) and $4 \times 4 \times 6$ (bottom).}\label{fig5}
\end{figure}

To analyze the density more closely, we divide the area of $ - 20 \le {\mathop{\rm Re}\nolimits} \left(2 \beta h \right) \le 20$, $ 0 \le {\mathop{\rm Im}\nolimits} \left(2 \beta h \right) \le \pi$ (two dimensions) or $ - 30 \le {\mathop{\rm Re}\nolimits} \left(2 \beta h \right) \le 30$, $ 0 \le {\mathop{\rm Im}\nolimits} \left(2 \beta h \right) \le \pi$ (three dimensions) into $400 \times 200$ boxes and count the number of zeros in each box. Figures \ref{fig4} and \ref{fig5} show the resulting density plots in two and three dimensions, respectively. It is readily seen that the density is very high on (and around) the imaginary axis in all cases. Regions of low density are cleared to white as can be verified by comparison of figure \ref{fig2} (left-bottom panel) and figure \ref{fig4} (bottom panel), for example. Zeros in those regions are expected to have no influence on the system properties in the thermodynamic limit. A marked difference between two and three dimensions is that the former develops a very sharp wedge in the body of the density profile as the size increases whereas, in the latter, the body remains rounded at its bottom part as commented above already. In two dimensions, the edge approaches the origin while the body does not show such an approach to the real axis at a faster rate than the edge with size increase. Also in three dimensions, the body seems to approach the real axis, but the difference between two cases in figure \ref{fig5} is not clear since the difference of size is small.

\section{Approach to the real axis}

Based on the qualitative observation in the previous section, we analyze our data quantitatively in this section. We assume that the behaviour of the nearest zero to the real axis of each sample determines the phase transition of the $\pm J$ model in the thermodynamic limit. We therefore calculate the average location of the nearest zero of each sample and analyze the system-size dependence of this average location. For two dimensions, we choose the temperatures as $T=T_c^{(\mathrm{pure})} = 2/\log ( \sqrt{2} +1 )$ $( \simeq 2.269 )$ and $T=0.5$ (in the presumed Griffiths phase). In three dimensions, the investigated temperatures are $T=T_g=1.1$ (the spin glass transition temperature \cite{kawashima}) and $T=0.5$ (in the spin glass phase).

\begin{figure}[tbp]
\begin{minipage}[t]{.5\textwidth}
\begin{center}
\includegraphics[width=1.0\hsize]{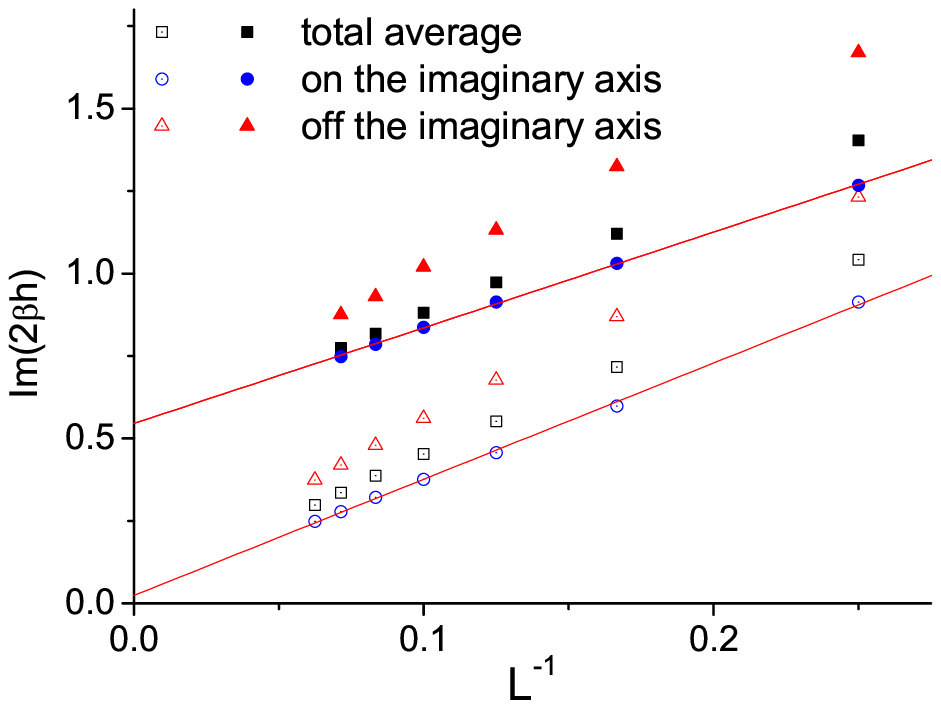}
\end{center}
\end{minipage}
\begin{minipage}[t]{.5\textwidth}
\begin{center}
\includegraphics[width=1.0\hsize]{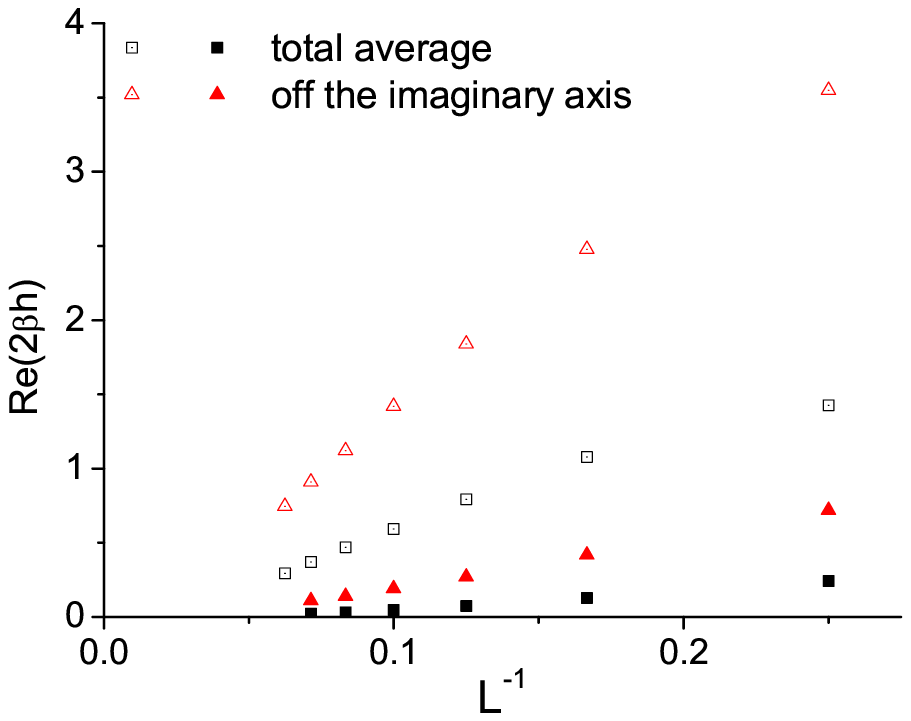}
\end{center}
\end{minipage}
\caption{Average locations of nearest zeros as functions of the inverse of the linear size in two dimensions for $T=T_c^{(\mathrm{pure})}$ (filled symbols) and $T=0.5$ (open symbols). The averages are calculated for zeros on the imaginary axis, off the imaginary axis, and both of these taken into account. Solid lines represent the best fits to linear functions.}
\label{fig6}

\begin{minipage}[t]{.5\textwidth}
\begin{center}
\includegraphics[width=1.0\hsize]{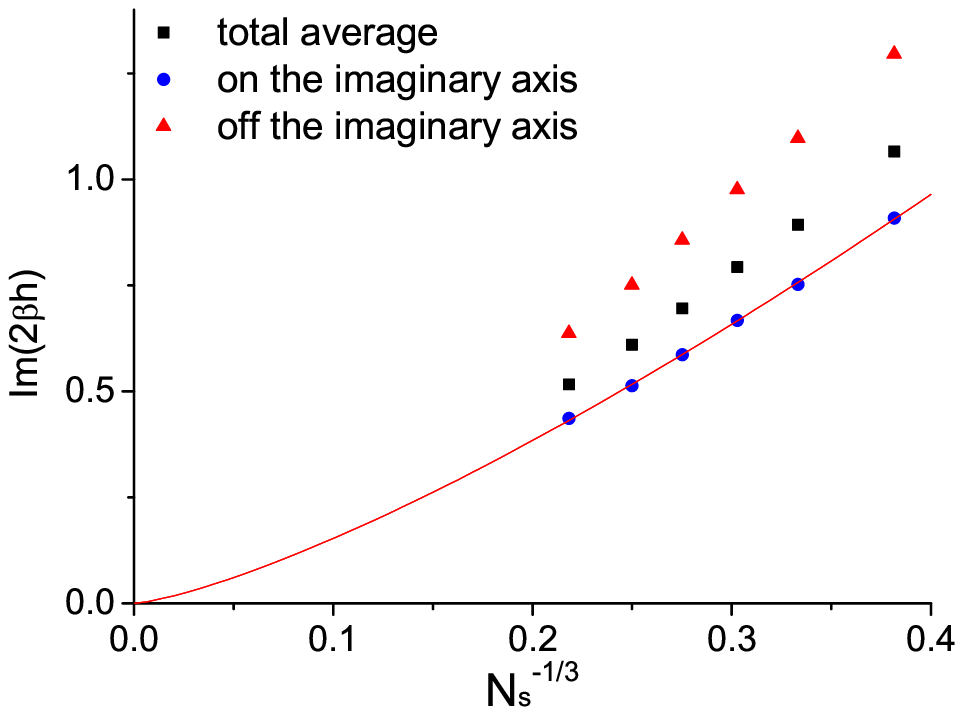}
\end{center}
\end{minipage}
\begin{minipage}[t]{.5\textwidth}
\begin{center}
\includegraphics[width=1.0\hsize]{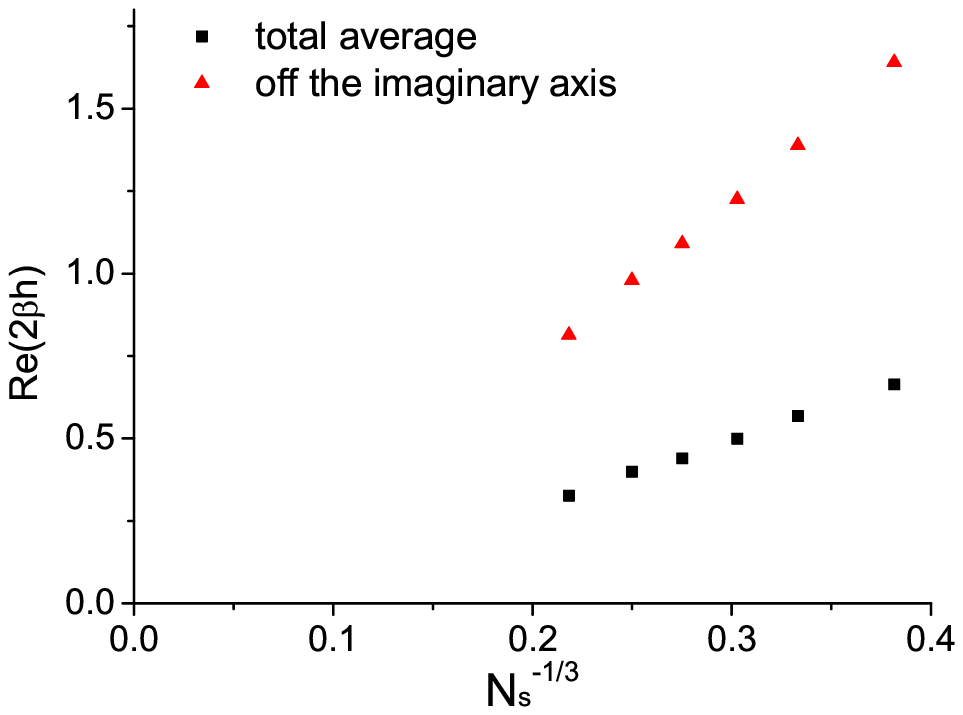}
\end{center}
\end{minipage}
\begin{minipage}[t]{.5\textwidth}
\begin{center}
\includegraphics[width=1.0\hsize]{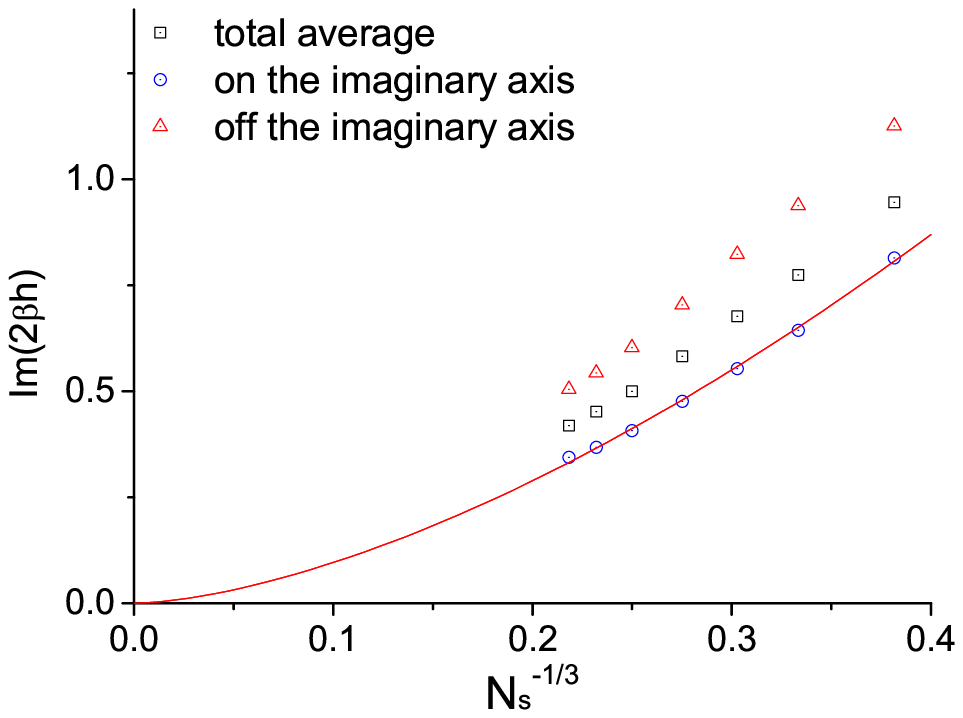}
\end{center}
\end{minipage}
\begin{minipage}[t]{.5\textwidth}
\begin{center}
\includegraphics[width=1.0\hsize]{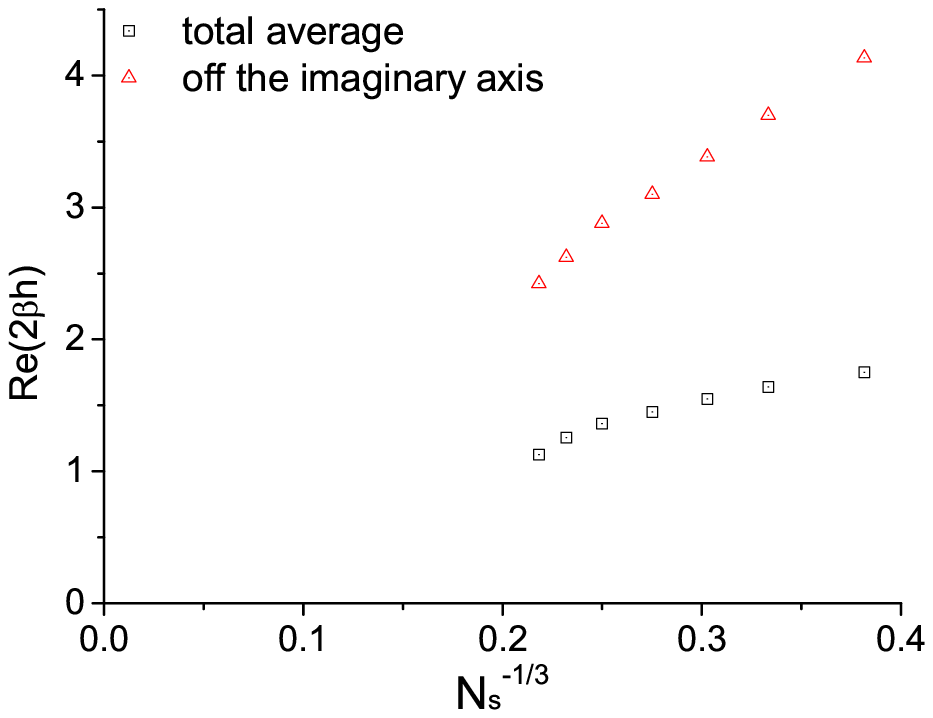}
\end{center}
\end{minipage}
\caption{Average locations of nearest zeros as functions of the inverse of the linear size in three dimensions for $T=T_g=1.1$ (top figures) corresponding to the spin glass transition temperature and $T=0.5$ (bottom figures). Solid lines represent best fits to power functions, $L^{-1.33}$ at $T=1.1$ and $L^{-1.59}$ at $T=0.5$.}
\label{fig7}
\end{figure}

Figures \ref{fig6} and \ref{fig7} show the real part (right part of figures) and the imaginary part (left part of figures) of the average location of nearest zeros as functions of the inverse of the linear size $L(=N_s ^ {1/d} )$ in two and three dimensions \footnote{See the Appendix for reasons to choose $N_s^{1/3}$ as the linear size in three dimensions.}.

We find that the average location of nearest zeros on the imaginary axis is lower than those off the imaginary axis for both dimensions as has been seen in previous figures. We may safely conclude that, if there is a phase transition caused by an approach of zeros to the origin, zeros on the imaginary axis reach the origin before those off the imaginary axis. 

In two dimensions, the imaginary part of the average location seems to approach the origin linearly with the inverse of system size. Solid lines in figure \ref{fig6} show the best fits to linear functions. The average location has clearly a non-vanishing imaginary part in the limit $L \rightarrow \infty$ for $T=T_c^{(\mathrm{pure})}$. The data for $T=0.5$ is marginal: It is difficult to discern whether or not the average location reaches the origin in the thermodynamic limit. Analysis from a different standpoint will be presented in the next section. The real part of the average location of zeros off the imaginary axis seems to rapidly approach the origin as $L \rightarrow \infty$.

In three dimensions, we tried the fit of the points on the imaginary axis to $L^{-\mu}$, because we expect the zeros to reach the origin in the thermodynamic limit due to the existence of the spin glass phase. Solid lines on the left parts in figure \ref{fig7} represent the best fits to the power function, $L^{-1.33(1)}$ for $T=T_g=1.1$ and $L^{-1.59(3)}$ for $T=0.5$. 

The value $\mu = 1.33(1)$ for the critical point $T=T_g=1.1$ in three dimensions may not be readily identified with a critical exponent: In the pure system it is known that the edge approaches the origin as $L^{-y_h}$ \cite{Creskim}. However, in the present random system, we evaluated the {\it average} location of the nearest zeros, for which we have no established scaling that relates $\mu=1.33(1)$ with critical exponents. In two dimensions our system does not show conventional critical behaviour at $T=T_c^{(\mathrm{pure})}$, and the exponent $\mu = 1$ in figure \ref{fig6} would not have direct relevance to critical exponents.

It is not easy to draw a definite conclusion on the behaviour of zeros in the thermodynamic limit in three dimensions since the system size is small. Nevertheless the following story is not inconsistent with our data: At $T=T_g$, if the average location of zeros reaches the real axis, it is likely to be only at the origin. At low temperatures, some people may wish to interpret the data that the system-size dependence of average zeros off the imaginary axis suggests the possible stability of the spin glass phase in the external field. If the imaginary part of the average of zeros off the imaginary axis approaches $0$ (triangles on the left bottom of figure \ref{fig7}) and the real part approaches a finite value (triangles on the right bottom), the spin glass phase would be stable in the external field. However, the data on the right-bottom panel of figure \ref{fig7} may instead be extrapolated to $0$, rather than to a finite value, in the thermodynamic limit. If this is indeed the case, it might imply the absence of the AT line \cite{AT} in the three dimensional $\pm J$ Ising model. We should anyway be very careful to draw a conclusion from the present data for small systems.

\section{Density of zeros on the imaginary axis}

In order to analyze the behaviour of zeros in a more detailed way, we study the density of zeros on the imaginary axis in two dimensions. We observe that zeros on the imaginary axis are in close proximity to the real axis in figures \ref{fig1} to \ref{fig5}. In particular we are interested in whether or not the density shows an essential singularity at the origin as an indicator of the Griffiths singularity. The density of zeros on the imaginary axis will be denoted as $g \left( \theta \right)$, where $\theta$ is defined by $2 \beta h = i \theta$ for the location of the edge, the nearest zero to the origin on the imaginary axis, of each sample. In order to compare the density of the $\pm J$ model with that of the diluted ferromagnet, we also estimated the density for the bond-diluted ferromagnet with the bond probability being 1/2 at $T=1.0$, which is believed to show an essential singularity at the origin \cite{Bray}.

\begin{figure}[tbp]
\begin{minipage}[t]{\textwidth}
\begin{center}
\includegraphics[width=0.5\hsize]{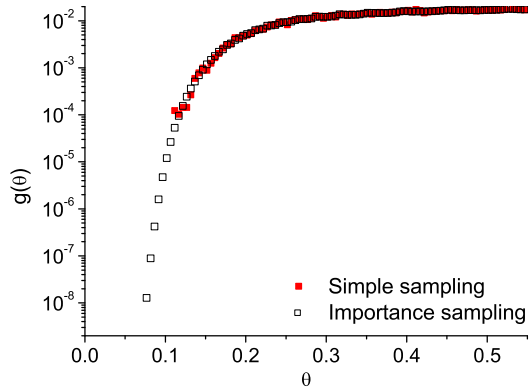}
\caption{Density of the nearest zero at $T=0.5$ for the system size $8 \times 8$ in the two-dimensional $\pm J$ model. The open squares have been obtained by the importance sampling MC algorithm, and filled squares are by the simple sampling of $10^5$ samples.}
\label{fig8}
\end{center}
\end{minipage}
\end{figure}

The density of nearest zeros falls very rapidly as shown in figure \ref{fig8} both for the conventional simple sampling and for the importance sampling (to be explained now). Since we are interested in the behaviour of $g \left( \theta \right)$ where its value is extremely small, a special care should be taken to reduce statistical errors in the estimate of $g \left( \theta \right)$. In general, the evaluation of events that occur with exponentially small probabilities is very difficult if one employs a simple-sampling procedure. We therefore used the importance-sampling Monte Carlo algorithm as follows \cite{huku, Hartmann, Korner, Monthus, HandI}.

The basic idea is to generate bond configurations according to the Metropolis algorithm. The new bond configuration $\mbox{\boldmath $J$}'$ is generated from the current configuration $\mbox{\boldmath $J$}$ with the probability
\begin{equation}
\min \left( {\frac{{\tilde{P}\left( {\theta \left( \mbox{\boldmath $J$} \right) } \right)}}{{\tilde{P}\left( {\theta \left( \mbox{\boldmath $J$}' \right) } \right)}},1} \right),
\label{MC}\end{equation}
where $\theta \left( \mbox{\boldmath $J$} \right)$ is the location of the edge of a given bond configuration $\mbox{\boldmath $J$}$ and $\tilde{P} \left( \theta \right)$ is a guiding function. The guiding function is ideally equal to $g \left( \theta \right)$ because equation (\ref{MC}) then guarantees that the stationary distribution of the Markov chain with equation (\ref{MC}) becomes the inverse of the guiding function, $1/\tilde{P}\left( \theta  \right)$. Therefore, configurations with smaller $g \left( \theta \right)$ are generated with higher probabilities. The idea is that the new bond configuration with a small probability $\tilde{P}\left( {\theta \left( \mbox{\boldmath $J$} \right) } \right)$ is encouraged to be chosen. In order to proceed without the prior knowledge of $g \left( \theta \right)$, we first choose an appropriate function as $\tilde{P}\left( {\theta \left( \mbox{\boldmath $J$} \right) } \right)$ and try an incremental improvement as follows.

After an update, the nearest zero $\theta \left( \mbox{\boldmath $J$}' \right)$ is calculated from the new bond configuration $\mbox{\boldmath $J$}'$, and the histogram $H \left( \theta \right)$ is incremented at this $\theta$ as $H \left( \theta \right) := H \left( \theta \right) +1$. Then the histogram $H \left( \theta \right)$ will reach the stationary distribution, $1/\tilde{P}\left( \theta  \right)$ multiplied by the true distribution $g\left( \theta  \right)$,
\begin{equation}
H\left( \theta  \right) \propto g\left( \theta  \right) \times \frac{1}{{\tilde{P}\left( \theta  \right)}}.
\end{equation}
Thus, we can estimate $g\left( \theta  \right)$ as
\begin{equation}
g\left( \theta  \right) \propto H\left( \theta  \right) \times \tilde{P}\left( \theta  \right).
\end{equation}
In our calculations, the new bond configuration $\mbox{\boldmath $J$}'$ is generated by flipping a single bond out of the current bond configuration $\mbox{\boldmath $J$}$. The initial guiding function $\tilde{P}\left( \theta  \right)$ is generated from the simple-sampling algorithm for about $10^5$ independent bond configurations. We repeat the above process to update $\tilde{P}\left( \theta  \right): = H\left( \theta  \right) \times \tilde{P}\left( \theta  \right)$ until $H \left( \theta \right)$ becomes nearly flat for the desired range, and finally we suppose that $\tilde{P}\left( {\theta} \right)$ is close enough to $g\left( {\theta} \right)$. The estimated autocorrelation time of our data is about a few Monte Carlo steps. However, we employ the data taken at every step because we observed no visible differences in the log scale by the change of steps to take data.

\begin{figure}[tbp]
\begin{minipage}[t]{.5\textwidth}
\begin{center}
\includegraphics[width=1.0\hsize]{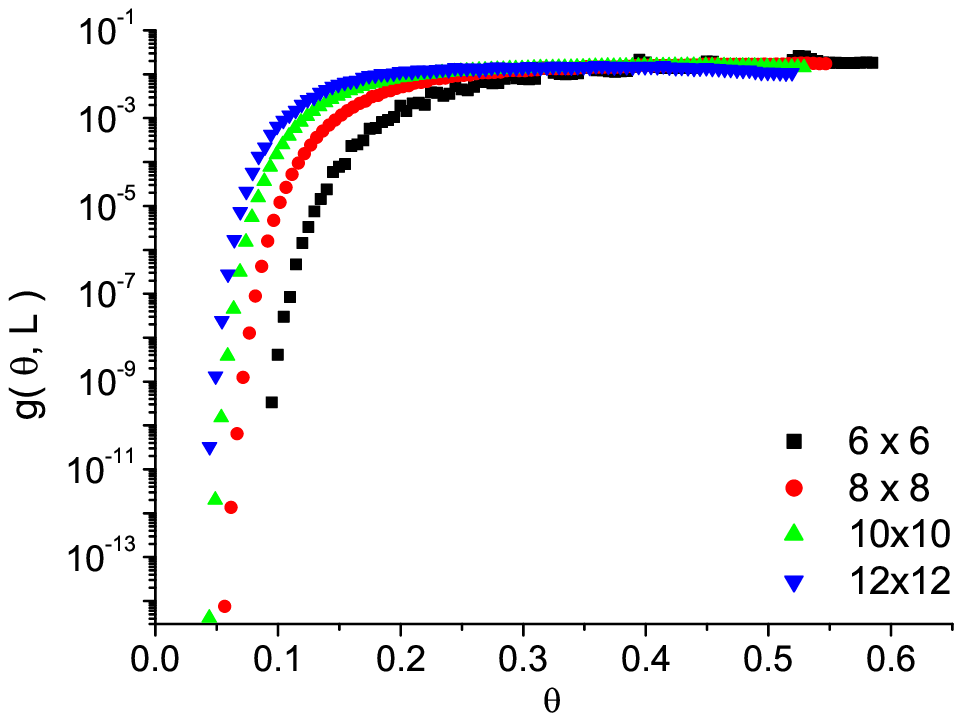}
\end{center}
\end{minipage}
\begin{minipage}[t]{.5\textwidth}
\begin{center}
\includegraphics[width=1.0\hsize]{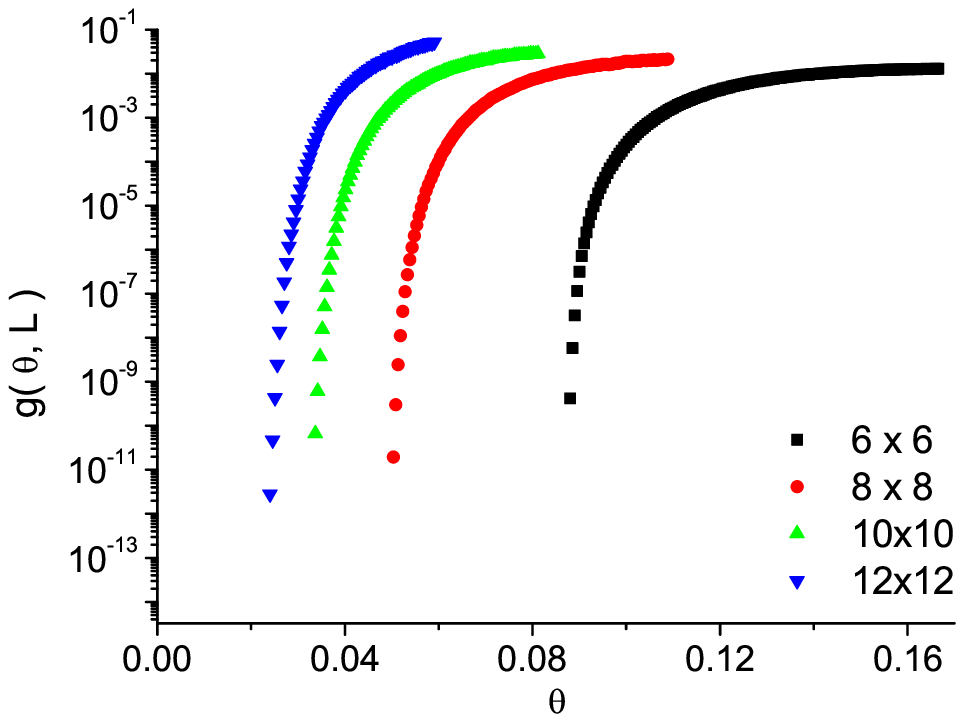}
\end{center}
\end{minipage}\caption{Density of the nearest zero for sizes $6 \times 6$ to $12 \times 12$ in two dimensions at $T=0.5$ for the $\pm J$ model (left) and $T=1.0$ for the diluted ferromagnet (right). }\label{fig9}
\end{figure}

\begin{figure}[tbp]
\begin{minipage}[t]{.5\textwidth}
\begin{center}
\includegraphics[width=1.0\hsize]{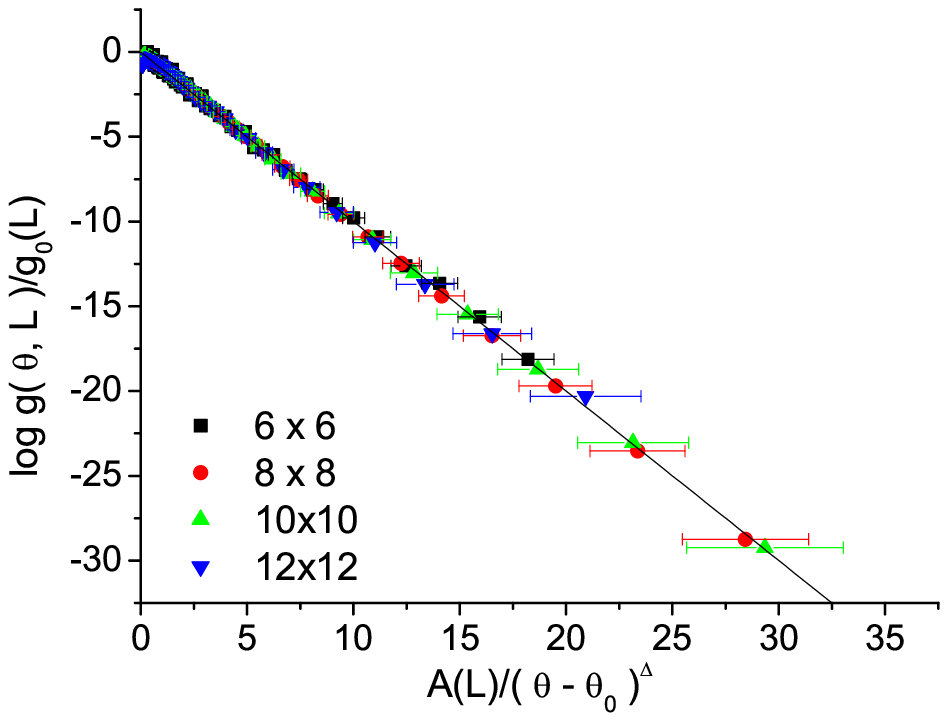}
\end{center}
\end{minipage}
\begin{minipage}[t]{.5\textwidth}
\begin{center}
\includegraphics[width=1.0\hsize]{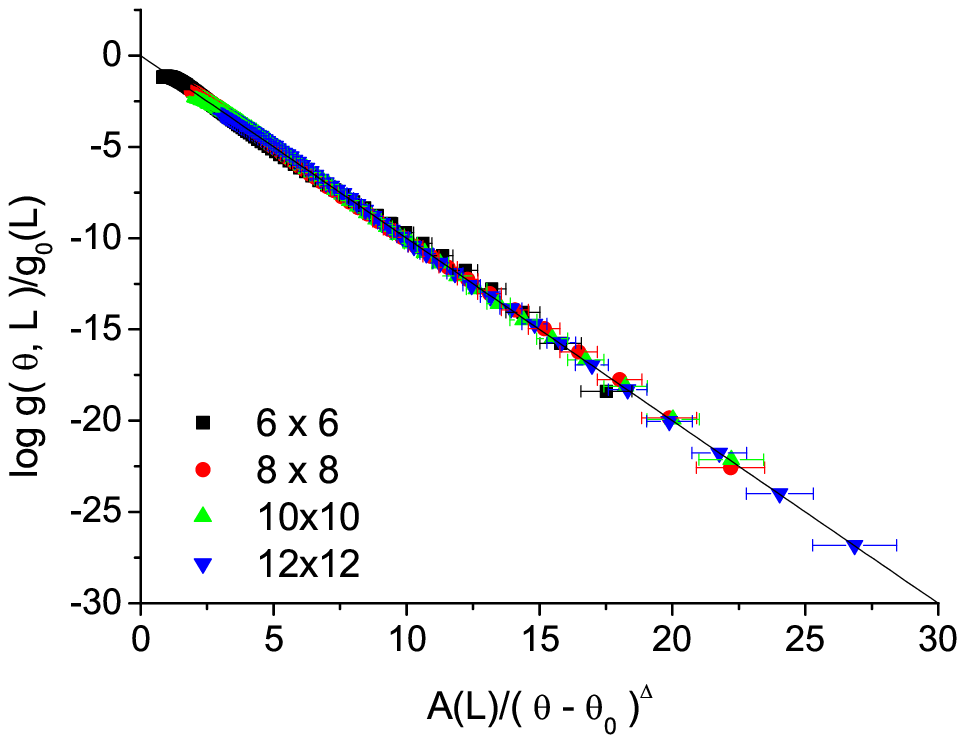}
\end{center}
\end{minipage}\caption{Best fits to the exponential function $g\left( {\theta ,L} \right) = g_0 \left( L \right) \exp \left[ { - A \left( L \right)/\left( \theta - \theta _0 \left( L \right) \right)^\Delta  } \right]$ with $\Delta = 2.2(1)$ for the $\pm J$ model (left) and $\Delta = 1.0(1)$ for the diluted ferromagnet (right), where $g_0 \left( L \right)$ denotes the parameter for normalization.
}\label{fig10}
\end{figure}

We have obtained the data for $g \left( \theta \right)$ of the $\pm J$ model and the diluted ferromagnet. Figure \ref{fig9} shows the estimated density of the edge for $L=6, 8, 10, 12$ in two dimensions at $T=0.5$ for the $\pm J$ model and $T=1.0$ for the diluted ferromagnet. These temperatures are below the transition points of the corresponding pure ferromagnets and therefore the systems lie in the Griffiths phase if any. With an essential singularity at the origin in mind, we analyze the data by the following function in consideration of the finite-size effect
\begin{equation}
g\left( {\theta ,L} \right) \approx \exp \left[ { - \frac{{A\left( L \right)}}{{\left( {\theta  - \theta _0 \left( L \right)} \right)^\Delta  }}} \right],
\label{finiteg}\end{equation}
where $\theta _0 \left( L \right)$ is an adjustable parameter expected to vanish in thermodynamic limit \footnote{A fit to a power, $g \left( \theta \right) \propto \left( \theta - \theta _0 \left( L \right) \right) ^ {- \Delta}$, gave $\theta$-dependent $\Delta$, an inconsistent result.}. The best fits to equation (\ref{finiteg}) have been obtained by choosing $ \Delta = 2.2(1)$ for the $\pm J$ model and $ \Delta = 1.0(1)$ for the diluted ferromagnet as shown in figure \ref{fig10} \footnote{Error estimates in these exponents are chosen to be on the safe side such that the error bars in figure \ref{fig10} are well included by the margin of several times.}. For the infinite system, $\theta _0 \left( L \right)$ is observed to vanish and the density becomes
\begin{equation}
g\left( \theta  \right) \approx \exp \left[ -A/\theta^{\Delta} \right].
\end{equation}
This function has an essential singularity at $\theta = 0$.  If $\Delta = 1$, it is the same function as suggested by Bray and Huifang for the diluted ferromagnet \cite{Bray}. It is remarkable that the analytical result of \cite{Bray} has been confirmed numerically and, in addition, a similar result has been observed for the $\pm J$ model. This fact suggests that a common physics is likely to underlie the behaviour, which supports the existence of the Griffiths phase also in the $\pm J$ model although connected clusters are not trivially defined in the spin glass system.

\section{Conclusion}

In this paper we have evaluated the zeros of the partition function of the symmetric $\pm J$ model in two and three dimensions by the transfer matrix method. The results revealed several outstanding features of the distribution of zeros in the complex field plane.

On the qualitative aspects, we found that the distribution has a real part on the complex field plane.  Nearest zeros to the origin lie on the imaginary axis. This edge moves toward the real axis as the system size increases in both dimensions. For the problem of the existence of the Griffiths singularity, we estimated the density of the nearest zero on the imaginary axis for each sample using the importance-sampling Monte Carlo algorithm. The idea of this algorithm is to enhance the probability of events with low probabilities. As a consequence, we could observe the density $g \left( \theta \right) \approx \exp ( -A/\theta ^ \Delta)$ for both the $\pm J$ model and the diluted ferromagnet for very small $\theta$ and extremely small $g\left( \theta \right)$. This function has an essential singularity at $\theta =0$: Thus this result for the density of the $\pm J$ model is compatible with the existence of the Griffiths singularity similarly to the diluted ferromagnet albeit through a slightly different mechanism as indicated by the difference between $\Delta = 2.2(1)$ for the $\pm J$ model and $\Delta = 1.0(1)$ for the diluted ferromagnet. This is the first evidence for a Griffiths singularity in spin glass systems in equilibrium. Further work is in progress to clarify the temperature dependence of the density and other properties using the method of \cite{Janke} for example, which may give us useful information on the phase transition.

\ack

We would like to thank Y. Ozeki and H. G. Katzgraber for helpful comments. This work was partially supported by CREST, JST and by the Grant-in-Aid for Scientific Research on the Priority Area ``Deepening and Expansion of Statistical Mechanical Informatics" by the Ministry of Education, Culture, Sports, Science and Technology. Parts of the numerical calculations have been performed on the TSUBAME Grid Cluster at the GSIC, Tokyo Institute of Technology.

\appendix
\section{The nearest zero of the pure system}

We can confirm the precision of our calculations from the computation of zeros of pure systems with uniform $J_{ij}=J>0$, for which all zeros are on the imaginary axis by the circle theorem. It is important to check the precision in these calculations, because we use the canonical transfer matrix which treats $x$ as numerical values in the partition function $Z \left( x,y \right)$. In tables \ref{tbl1} and \ref{tbl2} we list the location of the edge, nearest zero, as functions of the system size in two and three dimensions, respectively. These results for $L \ge 17$ have been obtained by the method of linear equations.
\begin{table}[btp]
\begin{tabular}{c|c|c}
\hline
$L$ & $\theta \left( T=T_c^{(\mathrm{pure})} \right)$ &$\theta \left( T=0.5 \right)$\\
\hline\hline
3 & 0.471998042059183  & 0.349071143945745\\
4 & 0.281611162550514  & 0.196350840622360\\
5 & 0.188399320772220 & 0.125664363108417\\
6 & 0.135505282010452 & 0.087266846351077\\
7 & 0.102460660627085 & 0.064114379665195\\
8 & 0.080370500990585 & 0.049087550024987\\
9 & 0.064840294163848 & 0.038785211232307\\
10 & 0.053487453492100  & 0.031416012357978\\
11 & 0.044925672320366  & 0.025963640664511\\
12 & 0.038302191885141  & 0.021816666139571\\
13 & 0.033068293166000  & 0.018589345677217\\
14 & 0.028857493586900  & 0.016028566259425\\
15 & 0.025417226074202  & 0.013962660497322\\
16 & 0.022568705664921  & 0.012271868296090\\
17 & 0.020182412161811  & 0.010870581295119\\
18 & 0.018162599374508  & 0.009696289311544\\
19 & 0.016437216606020  & 0.008702486168077\\
20 & 0.014951180535381  & 0.007853993248585\\
\hline
\end{tabular}
\caption{Locations of the edge of the pure system in two dimensions for $L=3$ to $20$ at $T=T_c^{(\mathrm{pure})}$ (left) and $T=0.5$ (right).}
\label{tbl1}
\end{table}

\begin{table}[btp]
\begin{tabular}{c|c|c}
\hline
$N_s$ & $\theta \left( T=T_c^{(\mathrm{pure})} \right)$ & $\theta \left( T=0.5 \right)$ \\
\hline\hline
$3\times 3 \times 2$ & 0.378263345499924   & 0.174532925915285   \\
$3\times 3 \times 3$ & 0.261003417586579    & 0.116355283788265    \\
$3\times 3 \times 4$ & 0.205196755580144    & 0.087266462782653    \\
$3\times 4 \times 4$ & 0.166725160223602    & 0.065449847087066   \\
$3\times 4 \times 5$ & 0.138347107456545    & 0.052359877648501  \\
$4\times 4 \times 4$ & 0.133611326102681    & 0.049087385315331  \\
$4\times 4 \times 5$ & 0.108662540409281    & 0.039269908236389   \\
$4\times 4 \times 6$ & 0.092979491978941   & 0.032724923521501   \\
$4\times 5 \times 6$ & 0.078908669455350 
      & 0.026179938817203   \\
$4\times 5 \times 7$ & 0.069174127961650 
      & 0.022439947553280    \\
$4\times 5 \times 8$ & 0.062199985131278 
      & 0.019634954106283     \\
\hline
\end{tabular}
\caption{Locations of the edge of the pure system in three dimensions for $N_s=3\times3\times2$ to $4\times5\times8$ at $T=T_c^{(\mathrm{pure})}$ (left) and $T=0.5$ (right). We used $T_c^{(\mathrm{pure})}=4.51$ \cite{blote}.}
\label{tbl2}
\end{table}

\begin{figure}[tbp]
\begin{minipage}[t]{.5\textwidth}
\begin{center}
\includegraphics[width=1.0\hsize]{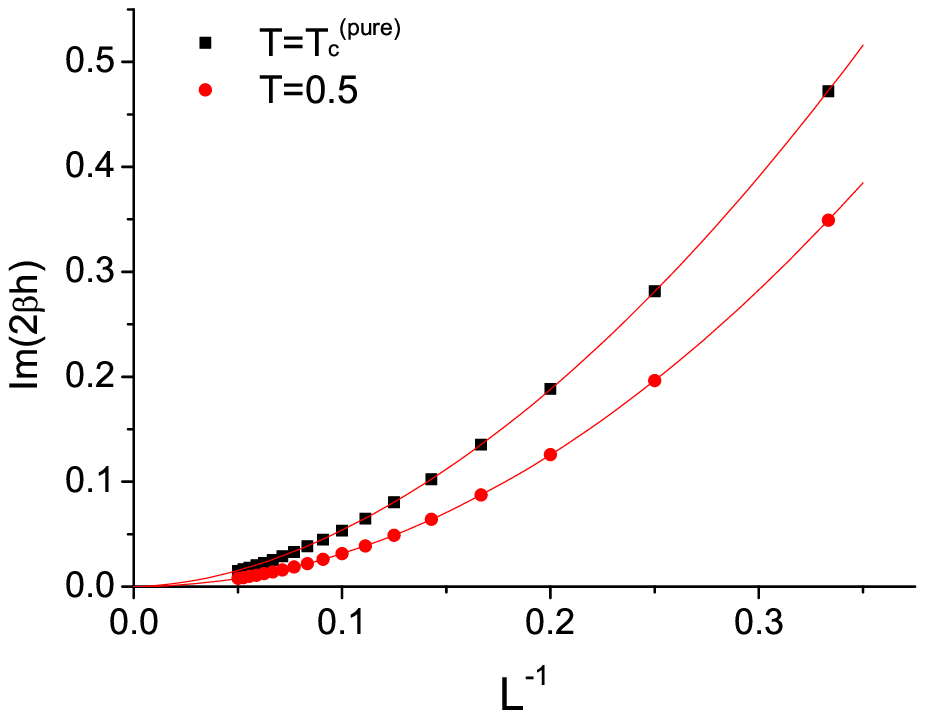}
\caption{Locations of the edge as a function of the system size $L^{-1}$. The solid line is the extrapolation by the BST algorithm.}
\label{figa1}
\end{center}
\end{minipage}
\begin{minipage}[t]{.5\textwidth}
\begin{center}
\includegraphics[width=1.0\hsize]{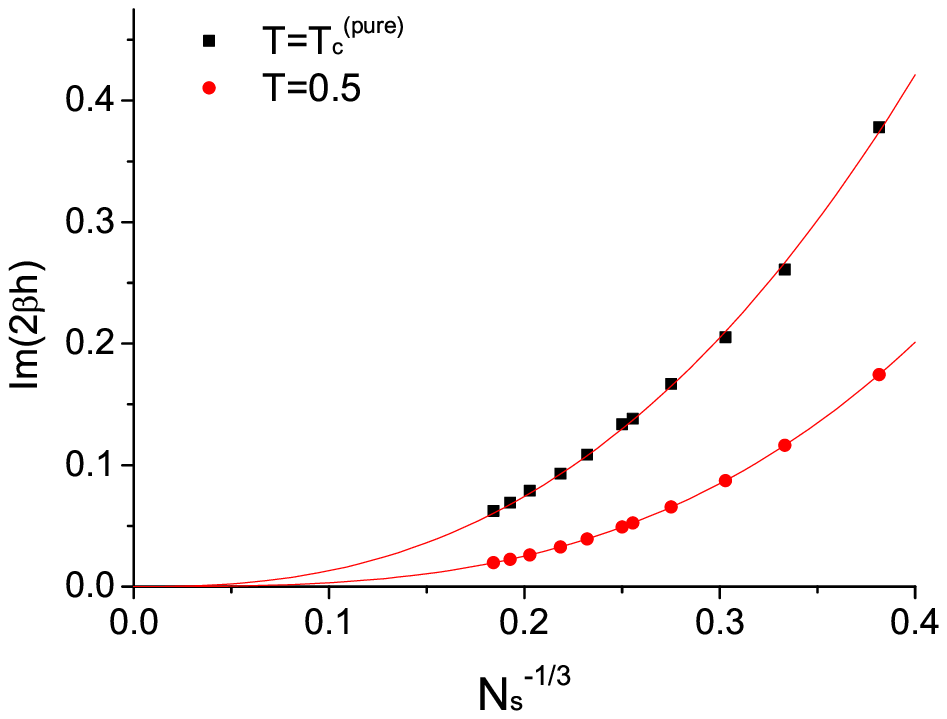}
\caption{Locations of the edge as a function of system size $N_s^{-1/3}$. The solid lines are fitting results as $\theta = C N_s^{-y_h/3}$.}
\label{figa2}
\end{center}
\end{minipage}
\end{figure}

To analyze the data, it is useful to remember that the angle $\theta$ is scaled as \cite{Creskim}
\begin{equation}
\theta =\cases{L^{ - y_h }&for $T=T_c^{(\mathrm{pure})}$\\
 L^{ - d}&for $T<T_c^{(\mathrm{pure})}$\\},
\label{A1}\end{equation}
where $y_h = 15/8$ $(d=2)$ and $y_h \simeq 2.48$ $(d=3)$ \cite{blote}. The location of the edge $\theta$ is expected to follow the scaling at $T=T_c^{(\mathrm{pure})}$ \cite{Creswick}
\begin{equation}
\theta \left( L \right) = CL^{ - y_h(L) } \left( {1 + C_1 L^{ - \omega }  + C_2 L^{ - 2\omega }  +  \cdots } \right),
\end{equation}
where $y_h(L)$ is defined as
\begin{equation}
y_h \left( L \right) =  - \frac{{\log \left[ {\theta \left( {L + 1} \right)/\theta \left( L \right)} \right]}}{{\log \left[ {\left( {L + 1} \right)/L} \right]}}.
\end{equation}
The exponent is also expanded as
\begin{equation}
y_h \left( L \right) = y_h \left( {1 + C'_1 L^{ - \omega }  + C'_2 L^{ - 2\omega }  +  \cdots } \right).
\end{equation}
We extrapolate $y_h \left( L \right)$ to $L \rightarrow \infty$ using the BST algorithm with $\omega = 1$ \cite{BST}. The data in table \ref{tbl1} yields $y_h = 1.87476$ $\left( T=T_c^{(\mathrm{pure})} \right)$ and $2.00000$ $\left( T=0.5 \right)$ in two dimensions (figure \ref{figa1}), very close to equation (\ref{A1}). In three dimensions, it is not trivial to define the linear size $L$, since the systems are not regular hexahedrons. Figure \ref{figa2} presents the analyses by the fitting function $\theta = C N_s^{-y_h/3}$. Then we obtained $y_h = 2.49(3)$ $\left( T=T_c^{(\mathrm{pure})} \right) $ and $3.00000$ $\left( T=0.5 \right)$. These results agree well with equation (\ref{A1}).

For further analysis we tried to fit the data at $T=0.5$ to $\theta = C L^{-d}$ to obtain $C=3.14163$ $(d=2)$ and $3.14159$ $(d=3)$. We thus conclude
\begin{equation}
\theta = \pi L^{-d} \qquad \left( T<T_c^{(\mathrm{pure})} \right) ,
\end{equation}
with negligibly small higher order corrections. These results give us justifications of our precision and of the use of $N_s^{1/3}$ as the linear size in three dimensions.

\end{document}